\documentclass[twocolumn]{aastex63}

\defcitealias{KD19}{KD19}
\received{May 6, 2021}
\revised{August 1, 2021}
\accepted{August 16, 2021}

\shorttitle{Inner Disk Water Enrichment}
\shortauthors{Kalyaan et al.}

\graphicspath{{./}{figures/}}

\begin{document}

\title{Linking Outer Disk Pebble Dynamics and Gaps to Inner Disk Water Enrichment}

\correspondingauthor{Anusha Kalyaan}
\email{a\_k363@txstate.edu}

\author[0000-0002-5067-1641]{Anusha Kalyaan}
\affiliation{Department of Physics,
Texas State University, 749 N Comanche St,
San Marcos, TX, USA}

\author{Paola Pinilla}
\affiliation{Max-Planck-Institut f\"{u}r Astronomie, K\"{o}nigstuhl 17, 69117, Heidelberg, Germany.}
\affiliation{Mullard Space Science Laboratory, University College London, Holmbury St Mary, Dorking, Surrey RH5 6NT, UK.}

\author{Sebastiaan Krijt}
\affiliation{School of Physics and Astronomy, University of Exeter, Stocker Road, Exeter EX4 4QL, UK}

\author{Gijs D. Mulders}
\affiliation{Facultad de Ingenier\'{i}a y Ciencias, Universidad Adolfo Ib\'{a}\~{n}ez, Av. Diagonal las Torres 2640, Pe\~{n}alol\'{e}n, Santiago, Chile}
\affiliation{Millennium Institute for Astrophysics, Chile}

\author{Andrea Banzatti}
\affiliation{Department of Physics, 
Texas State University, 749 N Comanche St,
San Marcos, TX, USA}


\begin{abstract}

Millimeter continuum imaging of protoplanetary disks reveals the distribution of solid particles and the presence of substructures (gaps and rings) beyond 5-10~au, while infrared (IR) spectra provide access to abundances of gaseous species at smaller disk radii. Building on recent observational findings of an anti-correlation between the inner disk water luminosity and outer dust disk radius, we aim here at investigating the dynamics of icy solids that drift from the outer disk and sublimate their ice inside the snow line, enriching the water vapor that is observed in the IR. We use a volatile-inclusive disk evolution model to explore a range of conditions (gap location, particle size, disk mass, and $\alpha$-viscosity) under which gaps in the outer disk efficiently block the inward drift of icy solids. We find that inner-disk vapor enrichment is highly sensitive to the location of a disk gap, yielding for each particle size a radial “sweet spot” that reduces the inner-disk vapor enrichment to a minimum. For pebbles of 1-10~mm in size, which carry the most mass, this sweet spot is at 7-15~au, suggesting that inner gaps may have a key role in reducing ice delivery to the inner disk and may not allow the formation of Earths and super-Earths. This highlights the importance of observationally determining the presence and properties of inner gaps in disks. Finally, we argue that the inner water vapor abundance can be used as a proxy for estimating the pebble drift efficiency and mass-flux entering the inner disk.
 
\end{abstract}


\section{Introduction} \label{sec:intro}

Protoplanetary disk observations in the last decade have dramatically increased our understanding of the formation and evolution of disks and of planet formation within them. From unprecedented high resolution observations from the Atacama Large Millimeter Array (ALMA), we know that dust disks are rich in structures such as gaps, rings and cavities \citep[e.g.,][]{andrews20, cieza2021, huang18a,long18,vdm18}. Features like rings and gaps may form via several mechanisms, such as planet-disk interactions \citep[][]{linpap79,rice2006,paardekooper2007,zhu12}, snow lines \citep[][]{SL88,rosjo13,banzatti15,oku16}, density enhancements at the outer edge of the deadzone \citep[][]{flock15,pinilla16}, zonal flows \citep[][]{baisto14,simon14,suriano18} and secular gravitational instabilities \citep[][]{TI14,TI16,tomi18}, whereas cavities may be produced by the presence of multiple giant planets \citep[][]{zhu2011,2018A&A...617A..44K}, as well as photoevaporative winds \citep[][]{ercolanopasc17} and disk winds \citep[][]{suzuki16}. Recent observations have also revealed the presence of two forming protoplanets within the cavity of the PDS\,70 disk \citep{2018A&A...617A..44K,2019NatAs...3..749H,isella19}. Collectively, these morphological features suggest that processes of planet formation are underway, and as seen in the case of PDS\,70, may even be harboring planets. 

High-resolution ALMA images of disks reveal the distribution and dynamics of solid particles (ranging from sub-mm sized dust, and mm-cm sized pebbles) in the outer disk ($\gtrsim$~5--10\,au). \footnote{In this study, we refer to solid particles $<$ 1 mm in size as dust, and $\ge$ 1 mm in size as pebbles.} As first predicted by theory \citep{weid77} and later confirmed by observations, we know that solid particles drift inwards, as we observe sharp edges in the dust continuum, expected from models of dust evolution and radial drift \citep{ba14}. We also see that the radial extension of dust/pebbles is smaller than that of gas in disks \citep[e.g.,][]{ans18,fac19,kurt21}, which is a sign of the inward drift of solids \citep[][]{trapman19,trapman20,rosotti19}. Moreover, we observe that solids accumulate in pressure bumps in the gas \citep{2012A&A...538A.114P}. As they drift inward towards regions of higher pressure, if particles encounter a pressure bump they flow toward the peak in gas pressure, reduce or stop their radial drift, and accumulate there. The existence of substructures in the largest disks suggests dust or pebble trapping is essential for maintaining large outer disk solid reservoirs and preventing the rapid depletion of solids into the star \citep[e.g.,][]{2012A&A...538A.114P,huang18a,long18,appelgren20}. Further evidence for dust trapping in action comes from studies that investigate pressure bumps and dust concentration within them \citep{pinilla2015, 2018ApJ...869L..46D,rosotti20}. Among solid particles, pebbles in particular have unique aerodynamic advantages; they are able to move quickly compared to smaller dust particles, and at the same time, are accreted efficiently via pebble accretion to form Earths, super-Earths and even cores of gas-giants \citep[e.g.][]{LJ14,Bitsch18,lamb19}.

While ALMA's angular resolution and sensitivity to resolve structures in the outer disk regions are unmatched by any other facility, it is rarely able to access the inner $< 5$\,au of the disk even in the closest star-forming regions at 120-150~pc \citep[][]{andrews16}. This innermost disk region is the location of terrestrial planet formation in the disk and holds key clues not only for formation timescales of Earth-like planets, but also the physical nebular conditions under which these planets form, which influence their atmospheric compositions, volatile content and their future habitability \citep[e.g.,][]{Oberg21,Vent20}. Given its higher temperature, the inner few astronomical units of disks emit rich infrared spectra, providing otherwise-inaccessible information on volatiles and their abundances  \citep[such as H$_2$O, CO, HCN; see e.g.][]{cn08,cn11,salyk08,salyk11,salyk19,pontoppidan10,banzatti17,banzatti20,walsh15}. Volatiles such as H$_2$O are generally present only as ice on solid particles in the outer colder disk beyond its snow line (located within a few astronomical units from the star), but is gaseous, abundant and observable via spectra of the inner disk where it sublimates out of these ice-bearing solids that drift inwards \citep[see][review]{2014prpl.conf..363P}. However, cold water vapor likely originating from photodesorption of water-ice has been detected in the outer disk with Herschel/HIFI data \citep{hog11}. Water vapor abundances in the inner disk have therefore long been expected to be linked to the mass of drifting icy pebbles that deliver water into the inner disk \citep[e.g.][]{cuzzi2004,cc06}. Constraining pebble mass fluxes into the inner disk via measurements of water abundances may thus uniquely inform models of rocky planet formation in the inner 3\,au region \citep{lamb19}.

Few studies have tapped into the synergistic advantages of combining information from both IR spectra and millimeter interferometry. \citet{2013ApJ...766..134N} first reported a correlation between flux ratio measurements of infrared HCN/H$_2$O with the dust disk mass as estimated from millimeter continuum fluxes from the Sub-millimeter Array, where disks with higher HCN/H$_2$O ratios appeared to have larger dust masses. The authors interpreted this correlation to be a result of efficient planetesimal formation in more massive disks where planetesimals accrete more solids, therefore locking water ice in the outer disk before it can reach the snow line and enrich the inner disk gas by sublimation. More recently, using ALMA disk images and a larger disk sample, \citet{banzatti20} studied correlations between the infrared luminosity of H$_2$O and of three carbon-bearing molecules (HCN, C$_2$H$_2$, and CO$_2$) and spatially-resolved dust disk radii, rather than disk masses that suffer from larger uncertainties \citep{aw05,bs91}. The analysis confirmed an anti-correlation between the H$_2$O/HCN ratio and dust disk radii, and expanded previous results by finding similar anti-correlations with H$_2$O/C$_2$H$_2$ and H$_2$O/CO$_2$ as well. After correcting for a common dependence on the accretion luminosity, this study further found that the strongest anti-correlation is between the water luminosity and dust disk radii, where small dusty disks ($\lesssim$60\,au in size) have higher water luminosity as compared to disks that are larger in the dust continuum emission (60 - 300\,au in size) and have substructures, either as gaps/rings, or an inner disk cavity, or both. \citet{banzatti20} interpreted this finding as supporting early predictions that the inner disk water vapor is enriched by the drift of icy pebbles from the outer disk that sublimate after crossing the snow line \citep[][]{cc06}, an enrichment that would be reduced in large disks with gaps and rings where the drift of icy pebbles is less efficient.

In this work, we test the interpretation of these observed trends by studying the influence of gaps in the outer disk on the water abundance in the inner disk. We build on the volatile-inclusive disk evolution model from \citet{KD19}, by including a gap at different radii in the outer disk in order to study the resulting pebble dynamics, and compute volatile abundances in the inner disk as a function of time.  
Section \ref{sec:methods} describes the model, and Section \ref{sec:results} the results of our simulations. In Section \ref{sec:disc}, we present the insights gained from these simulations and in Section \ref{sec:conc} we present the main conclusions of this study.

\section{Methods} \label{sec:methods}

To test how the outer disk solid dynamics may be linked to the inner disk vapor abundance, we employ the volatile-inclusive disk evolution model of \citeauthor{KD19} (\citeyear{KD19}; hereafter \citetalias{KD19}), as illustrated in Figure 1. Our 1D disk model evolves a protoplanetary disk around a solar-like star through several million years, across a computational grid of 0.1 - 500 au, with 300 radial zones. To this, we add the radial transport of solid particles, by incorporating their radial drift, advection and diffusion. Since water is present as ice in solids, and as vapor in the inner disk within the snow line, we compute the pressure-temperature conditions where ice sublimates to vapor, and include the diffusion of vapor in the inner disk. 
The above transport processes altogether determine the distribution and evolution of water across the snow line as well as the entire disk (see \citet{2015ApJ...815..112K} and \citetalias{KD19} for more details). As we are specifically interested in the influence of substructures in the outer disk on the inner disk water abundance, we create a gap in the outer disk gas. At the outer edge of this gap, we expect solid particles to be trapped within the pressure bump. In this section, we highlight the important features of the disk evolution model and detail how we incorporate a gap as substructure to study the effect of outer disk dust and pebble dynamics on inner disk volatile enrichment. We list the main parameters used in our simulations in Table 1.
\begin{figure*}[htb!]
	\centering
	\includegraphics[scale=0.35]{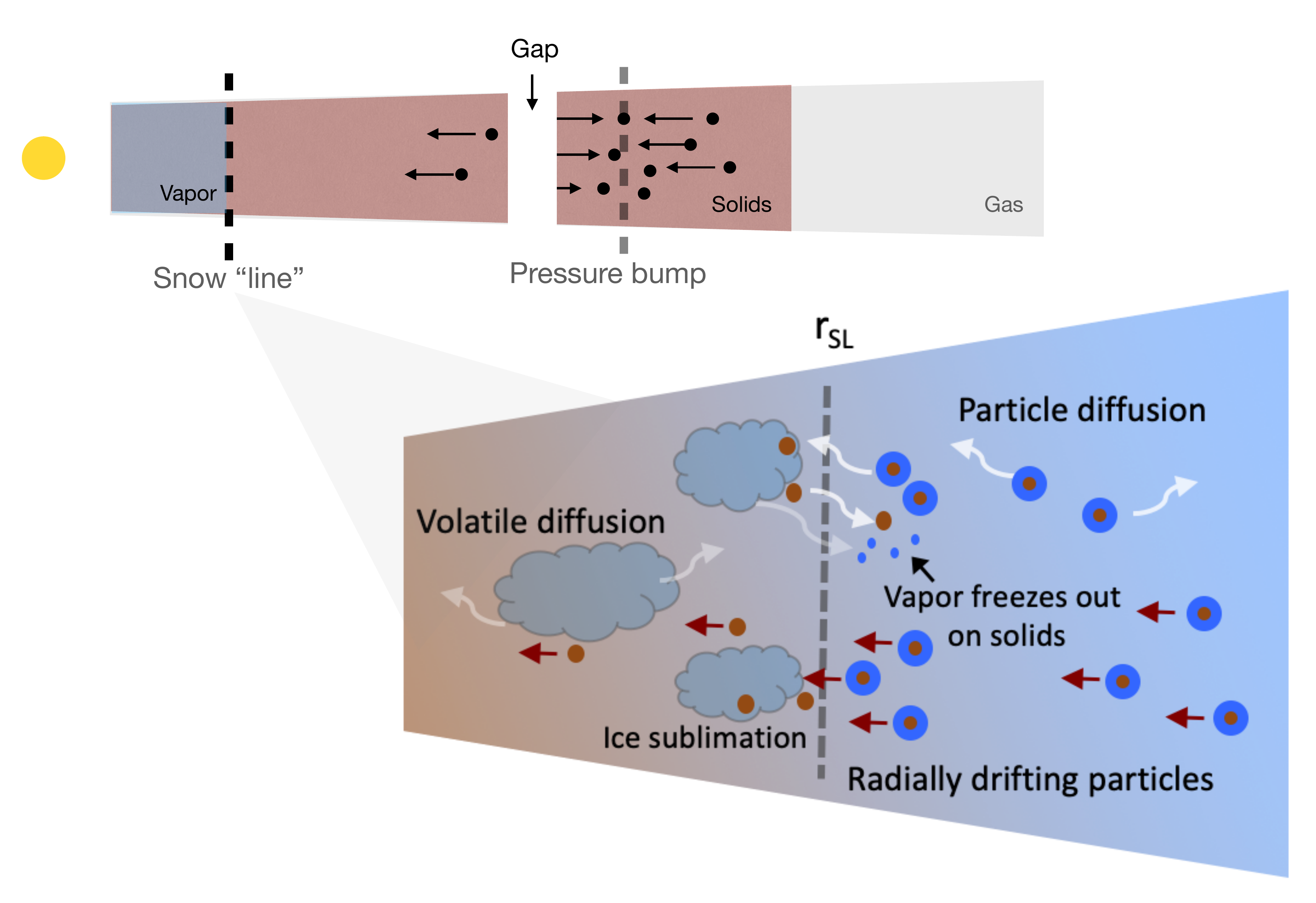}
	\caption{Schematic of our evolutionary disk model with solid and volatile transport across a disk with a gap. The processes depicted here follow a sequence: i) Solids radially drift inwards from the cold outer disk; small solid particles also diffuse throughout the disk. ii) On encountering a gap, they accumulate in the pressure bump just beyond the gap. iii) When these water-ice-bearing solids approach the snow line, ice in them sublimate and become vapor (blue plume), leaving behind bare silicate particles (brown circles). iv) Some vapor back-diffuses across the snow line to refreeze as ice on solids. v) Volatiles diffuse throughout in the inner disk. vi) Both solids and volatiles are eventually accreted onto the star. }
\end{figure*}

\subsection{Transport of Gas}

In our disk model, we use the standard equations of disk evolution from \citet{1974MNRAS.168..603L}, explicitly integrated in time, as below:
\begin{equation}
 \frac{\partial \Sigma} { \partial t} = \frac{1}{2 \pi r} \frac{\partial\, \dot{\rm M} }{\partial\, r} \,,
\end{equation}
where the rate of change of surface density $\Sigma$ is related to rate of mass flow through an annulus $\dot{\rm M}$ as:
\begin{equation}
  \dot{\rm M}=6 \pi r^{1/2} \frac{\partial}{\partial r} \left(r^{1/2}\Sigma \,\nu \right).
\end{equation}
Here, $\nu$ is the viscosity in the disk, assumed to have a turbulent origin, and defined with the following standard scaling relation \citep{1973A&A....24..337S}:
\begin{equation}
    \nu = \alpha \frac{c_s^2}{\Omega_k} \, ; 
\end{equation}
where $c_s$ is the local sound speed, $\Omega_k$ is the Keplerian angular frequency, and $\alpha$ parameterizes the efficiency of turbulent transport. We adopt a radially uniform $\alpha_{\rm const}$ = $10^{-3}$ as the canonical value in our simulations.

We assume the standard temperature profile of a passively-heated flared disk, as follows, where temperature T across mid-plane radius $r$ is assumed to be:
\begin{equation}\label{tempeqn1}
	{\rm T}(r) = \Big(\frac{\rm L_*}{4 \pi \sigma}\Big)^{1/4}\,\, \phi^{1/4}\,\, r^{-1/2}.
\end{equation}

Here, $\sigma$ is the Stefan-Boltzmann constant, $\phi$ is the incident angle parameter and  L$_*$ is assumed to be 0.8 L$_{\odot}$, the median luminosity among T-Tauri stars in \citet{banzatti20}, which yields:

\begin{equation}\label{tempeqn2}
{\rm T_{\rm pass}}(r) = 118 \, \left( \frac{r}{ 1 \, {\rm au}} \right)^{-0.5} \, {\rm K}.
\end{equation}

We neglect accretion heating in this work. (This assumption is discussed in Appendix \ref{app:accheat}. We find that our main conclusions remain unchanged with a change in the location of the snow line.)

\begin{table*}
\centering
\begin{tabular}{ |p{6cm}|p{3cm}|p{5cm}|  }
 \hline
 \bf{Model Parameter} & \bf{Symbol} & \bf{Parameter Value/Range}\\
 \hline
 
 \multicolumn{3}{|c|}{\bf{Disk Parameters}} \\
 \hline
 Inner disk radius   & R$_{\rm in}$    &    0.1 au\\
 Characteristic radius &   R$_{\rm char}$  & 70 au\\
 Outer disk radius &  R$_{\rm out}$ & 500 au\\
 Initial disk mass    & M$_{\rm disk}$ & 0.01, \textbf{0.05}, 0.1 M$_{\odot}$\\
 Disk viscosity& $\alpha_{\rm const}$ & 10$^{-4}$, \textbf{10$^{-3}$} \\
 \hline
 \multicolumn{3}{|c|}{\bf{Solid Particle Parameters}} \\
 \hline
 Solid particle size (diameter) & $a_{\rm p}$ & 0.1, 0.3, \textbf{1}, 3, 10 mm\\
 Internal solidparticle density & $\rho_{\rm p}$  & 2.5 g cm$^{-3}$\\
 Initial solid particle surface density & $\Sigma_{\rm icy solid}$, $\Sigma_{\rm rocky solid}$ & 0.005 $\times$ $\Sigma_{\rm gas}$ g cm$^{-2}$\\
 Schmidt number for diffusion & Sc$_{\rm p}$ & 0.7\\
 \hline
 \multicolumn{3}{|c|}{\bf{Vapor Parameters}} \\
 \hline
 Initial concentration of water vapor & $c$ & 1 $\times$ 10$^{-4}$\\
 Schmidt number for diffusion & Sc$_{\rm v}$ & 1.0\\
 \hline
 \multicolumn{3}{|c|}{\bf{Gap Parameters}} \\
 \hline
 Gap location & R$_{\rm gap}$ & 7, 15, \textbf{30}, 60 au\\
 Peak $\alpha$ at gap & $\alpha_{\rm gap}$ & 75 $\times$ $\alpha_{\rm const}$\\
 Time when gap forms & t$_{\rm gap}$ & 0.1 Myr\\
 \hline
\end{tabular}
\caption{Table of parameters used in our simulations. Bold values indicate fiducial model parameters.}
\end{table*}
\subsection{Transport of Solids}
To the underlying bulk gaseous disk, we add solid particles that range from sub-mm to cm sized. In the absence of gas, particles in the disk would maintain Keplerian orbital motion around the star. In the presence of gas, however, they lose angular momentum facing a headwind from the pressure-supported sub-Keplerian gas and therefore drift inwards. 

In this work, we keep track of two drifting populations of solids: rocky particles (composed of 100\% silicates), and icy particles (composed of 100\% ice). In all, the mass of the incoming solid population beyond the snow line is assumed to be 50\% ice and 50\% rock. This ice-to-rock ratio is generally consistent with cometary compostion in the solar system \citep[e.g.][]{mc11}.

These two populations are identical in size of particles, and thus identical in their transport, only differing in their composition. We adopt this treatment from \citetalias{KD19} to track the mass of water across the cold disk beyond the snow line and the change of phase of water as icy solids approach the snow line. We describe this in detail in Section \ref{sec:methods:water}. 

We consider solid transport in two drag regimes: Epstein and Stokes regimes \citep{weid77}, where the particle radius $a_{\rm p}$ may be smaller (Epstein regime) or larger (Stokes regime) than the mean free path $\lambda_{\rm mfp}$ at each $r$. This transition occurs at $a_{\rm p}$/$\lambda_{\rm mfp} = 9/4$ \citep{2010A&A...513A..79B}. We follow \citet[][]{krijt16} and use the following expression to evaluate the Stokes number St(r):

\begin{equation}\label{eqn:stokes}
{\rm St} = \frac{\pi}{2} \, \frac{\rho_{\rm p} a_{\rm p}}{\Sigma} \, \Big[ 1 + \frac{4}{9} \frac{a_{\rm p}}{\lambda_{\rm mfp}}\Big] , 
\end{equation}
where $\rho_{\rm p}$ = 2.5 g cm$^{\rm -3}$ is the assumed internal density of particles, equivalent to the density of silicates and $a_{\rm p}$ is the radius of the particle. $\lambda_{\rm mfp}$ = 1 / ($n \sigma_{\rm H_2}$), where $n$ is the mid-plane number density, and $\sigma_{\rm H_2}$ = 2.0 $\times$ 10$^{-15}$ cm$^{2}$. The above prescription allows for a smooth transition between the Epstein and the Stokes regimes. In practice, however, with the exception of the largest particles in the innermost disk at earliest times, our simulations operate entirely in the Epstein limit.
The drift velocity of particles with respect to the gas V$_{\rm drift}$ is defined as:
\begin{equation}\label{eqn:drift}
{\rm V_{\rm drift}} = \frac{ -{\rm St}^2 \, {\rm V_{\rm g,r}} - {\rm St} \, \eta \, r \,\Omega_k }{ 1 + {\rm St}^2 },
\end{equation}
where V$_{\rm g,r}$ is the radial velocity of the gas and $\eta$ = $-\, 0.5 (\partial \,\rm{ln}\, P/ \partial \,\rm{ln} \,r) \, c_s^2 / V_k^2$. Here, V$_{\rm k}$ is the Keplerian velocity.

As adopted in previous works (\citeauthor{2017ApJ...840...86D}, \citeyear{2017ApJ...840...86D}, \citeyear{2018ApJS..238...11D}; \citetalias{KD19}) the transport of particles is evolved with the following evolution equation that altogether considers advection (first term), drift (second term) and diffusion (third term):
\begin{equation}
\dot{\rm M}_{\rm p} = c_{\rm p} \dot{\rm M}\, - \,2 \pi r \,c_{\rm p}\, \Sigma\, {\rm V_{\rm drift}}\, + \,2\pi r \, \Sigma \, {\cal D}_{\rm p} \,  \frac{\partial \, c_{\rm p}}{\partial \,r},
\end{equation}
 where the concentration of particles to disk bulk gas is given by $c_{\rm p}$ = $\Sigma_{\rm p}/\Sigma$
and particle diffusivity ${\cal D}_{\rm p}$ is given by:
\begin{equation}\label{eqn:diff}
{\cal D}_{\rm p} = \frac{ {\cal D}_{\rm gas} }{ 1 + {\rm St}^2 }. 
\end{equation}
Here, ${\cal D}_{\rm gas}$ is the diffusivity of the bulk gas and is equivalent to turbulent viscosity $\nu$. We make the assumption that the $\alpha$ responsible for viscous disk evolution is also regulating particle and vapor diffusion. (Note that in equation 8, first and second terms combine to yield the total radial mass flux for drifting particles, with the total radial particle velocity being ${\rm V}_{\rm drift} + {\rm V}_{\rm g,r}$.) 

\subsection{Transport of Water in Vapor/Ice}\label{sec:methods:water}

We follow the same treatment used in \citetalias{KD19} in this work, similar to the treatment in \citet{cc06}. Depending on the pressure-temperature conditions in the disk, water is present either as ice in the outer colder disk (outside of the snow line) or as vapor in the warmer inner disk (inside of the snow line). To determine the local physical state of water, we use the following relations derived from experiments to calculate the local saturation vapor pressure over ice, i.e., the pressure where water vapor is in thermodynamic equilibrium with its condensed state, at any given temperature T, as follows: 
\begin{equation}
{\rm P_{\rm eq}(T)} = 0.1 \, \exp \left( 28.868 - 6132.9 / {\rm T} \right) \, {\rm dyn} \, {\rm cm}^{-2},  {\rm T} > 169 \, {\rm K}
\end{equation}
from \citet{1993GeoRL..20..363M}, and 
\begin{equation}
{\rm P_{\rm eq}(T)} = 0.1 \, \exp \left( 34.262 - 7044.0 / {\rm T} \right) \, {\rm dyn} \, {\rm cm}^{-2}, {\rm T} \leq 169 \, {\rm K}
\end{equation}
from \citet{2003GeoRL..30.1121M}.

We then compute the surface density of water vapor from the equilibrium vapor pressure using the following equation:
\begin{equation}
\Sigma_{\rm vap,eq} = (2\pi)^{1/2} \, \left( \frac{ {\rm P_{\rm eq}} }{ c_{\rm H_2O}^2 } \right) \, {\rm H},
\end{equation}
where H = $c_s / \Omega$ is the scale height of the disk, and $c_{\rm H_2O}$ 
the sound speed in water vapor.

If the total water content (in vapor and icy particles) at a given $r$ is less than the equilibrium vapor pressure, we assume all water is in vapor at that $r$.  Conversely, if the total water content (in vapor and icy particles) is greater than equilibrium vapor pressure, then we assume $\Sigma_{\rm vap} = \Sigma_{\rm vap,eq}$ and the rest of the water is in ice. In this work, we assume that freezing is instantaneous. 
The physical state of water in either form dictates how it will be transported. As mentioned before, we adopt two populations of solids, one made completely of ice, and one made completely of silicates. This approximates the presence of silicate particles covered by icy mantles. These two populations move identically, except when they approach the pressure-temperature conditions of the water snow line. When they drift into the snow line, the icy solids will sublimate to vapor as detailed above.

We adopt initial surface density of icy and rocky particles, i.e.,  $\Sigma_{\rm icypart}$ = $\Sigma_{\rm rockypart}$ = 0.005 $\times$ $\Sigma_{\rm gas}$, by using the standard solids-to-gas ratio of 0.01, multiplied by 0.5, i.e., half of the mass in solids in icy, and half in rocky particles. Therefore, for our fiducial model, at t=0, there are 166.5 Earth masses of solids across the entire disk, and $\sim$ 83 Earth masses each in rocky particles, and icy particles across the disk.

Within the snow line, transport of water in vapor form is assumed to follow the following equations from \citet{2017ApJ...840...86D} where vapor is treated as a diffusive tracer species in the bulk gas:
\begin{equation}
\frac{\partial \Sigma_{\rm vap}}{\partial t} = \frac{1}{2\pi r} \, \frac{\partial \dot{\rm M}_{\rm vap}}{\partial r},
\end{equation}
where the mass flux of vapor is
\begin{equation}
\dot{\rm M}_{\rm vap} = 2\pi r \, \Sigma \, {\cal D}_{\rm vap} \,  \frac{\partial c}{\partial r}.
\end{equation}
 Here, $c$ is the concentration of water vapor with respect to bulk disk gas, i.e., ($\Sigma_{\rm vap}$/$\Sigma$) and ${\cal D}_{\rm vap}$ = ${\cal D}_{\rm gas}$ $\approx$ $\nu$ (assuming Schmidt number Sc $\approx$ 1). (Some water vapor back-diffuses out of the snow line region. However, without a sink for planetesimal formation, this mass eventually makes its way back within the snow line. See \citetalias{KD19} for more details.)

\subsection{Formation of a Gap}

Since our motivation is to determine the effect of outer disk structure on water delivery to the inner disk, we include a gap to the disk structure by incorporating a perturbation to the $\alpha$ profile at the location of the gap. For each gap included in our simulations, we locally increase the $\alpha$ over the globally uniform $\alpha$ as a Gaussian profile (similar to previous works such as \citet[][]{2018ApJS..238...11D} and \citet[][]{stamm19}) as below:
\begin{equation}
    \alpha(r) = \alpha_{\rm const} + (\alpha_{\rm gap} - \alpha_{\rm const})\, \exp\,(-x^2) \,,
\end{equation}
where $\alpha_{\rm const}$ is the constant uniform value of $\alpha$ adopted throughout $r$ and $\alpha_{\rm gap}$ is the peak value assumed at the center of the gap = 75 $\times$ $\alpha_{\rm const}$. This sets the depth of the gap in the gas in our simulations and is generally consistent with hydrodynamic simulations of gaseous gaps created by a 1 M$_{\rm jup}$ planet \citep[see Figure 2, lower $h/r$, ][]{zhang18}. We keep this peak gap $\alpha$ the same irrespective of disk location to simulate a gap formation mechanism of similar strength at different $r$. Here, $x = (r - r_{\rm gap})$ / gap width, where the width of the gap is chosen to be 2 $\times$ H, i.e., twice the scale height of the disk. We choose this width as it is likely that only pressure bumps greater than 1 $\times$ H in width are long-lived and stable, and assume that the gap should be similar in width as the pressure bump \citep{ono16,2018ApJ...869L..46D}. We also assume that the gap in all simulations instantaneously opens at 0.1 Myr. 

\begin{figure*}[htp]
\centering
\includegraphics[scale=0.45]{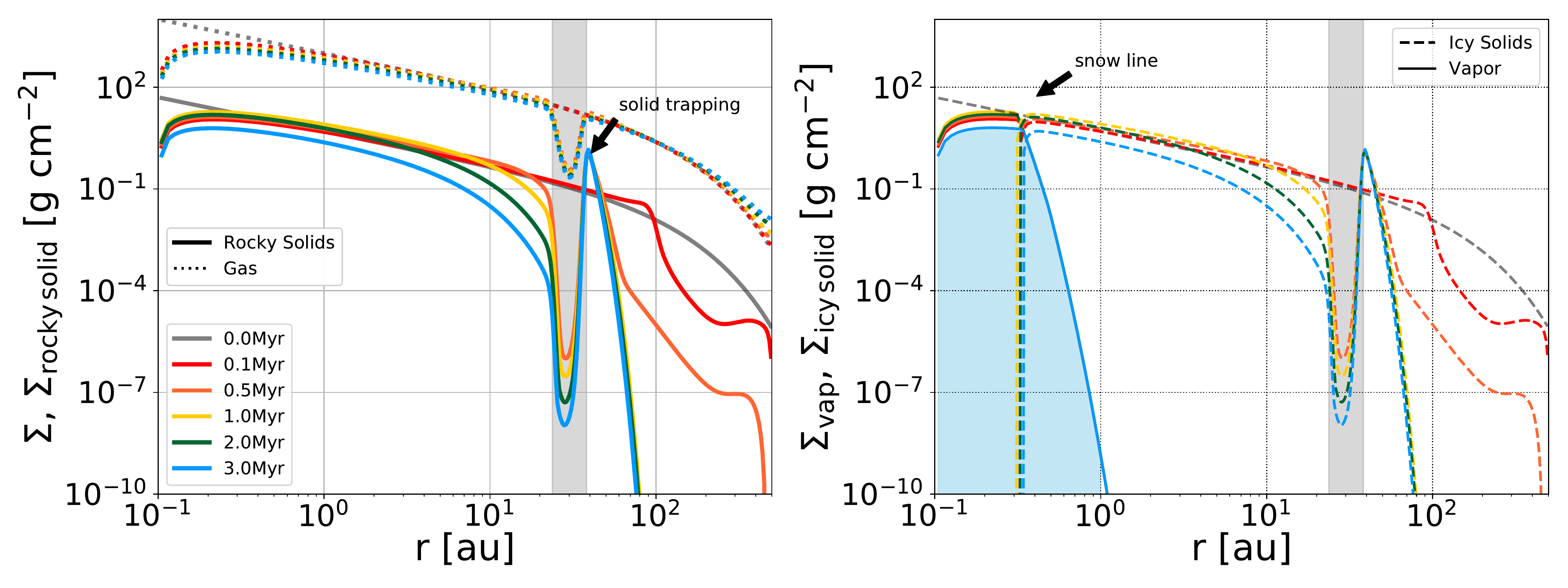}
\caption{Evolutionary profiles of surface densities of gas, solid particles (rocky and icy) and vapor for our fiducial model with particles of size 1 mm and disk mass of 5\% M$_{\odot}$ (see Table 1). Left panel shows the time evolution of surface density of gas $\Sigma$ (dotted) and of rocky solids $\Sigma_{\rm rocky\,solid}$ (solid), for a disk model with a gap at 30 au in gas. Particles are trapped in the pressure bump beyond the gap. Right panel shows the time evolution of $\Sigma_{\rm vapor}$ (thin solid, blue shaded area) and $\Sigma_{\rm icy\,solid}$ (dashed). Black, red, yellow, green and blue correspond to times t = 0, 0.1 (when gap opens), 0.5, 1, 2 and 3 Myr, respectively. Gray shaded area denotes region within the gap. A comparison of gaps at different radii is included in the Appendix.}
\label{fig:FIDU}
\end{figure*}

\section{Results} \label{sec:results}

We present in this section the results from the grid of simulations performed in this study. We first describe the results from the fiducial model, and then discuss the results from the rest of the grid exploring a range of gap radii, particle sizes, disk masses, and $\alpha$-viscosity.

\subsection{Our Fiducial Model}

Figure \ref{fig:FIDU} shows the evolutionary profiles of surface densities of the various components of our fiducial model: the underlying bulk gas disk, the populations of drifting solid particles as well as water vapor within the snow line region. In this fiducial model, we consider two  populations of mm-sized particles drifting inward (one composed of ice, the other of silicates) that encounter a gap in the gaseous disk at r$_{\rm gap}$ = 30\,au. Once the gap is opened, particles beyond it can be trapped at the pressure bump beyond 30\,au, as observed in the increase in $\Sigma_{\rm solid}$ beyond r$_{\rm gap}$ in the left panel. In all our runs, a gap is assumed to form instantaneously at 0.1 Myr (right after the solid red line in the left panel). Figure \ref{fig:FIDU} highlights the overall working of our model, where the left panel shows particle-trapping, while the right panel shows surface density of water in its various forms, i.e., as vapor within the snow line and as ice in drifting solids beyond the snow line ($>$0.3 au). As both icy and rocky solid populations are dynamically identical, they show identical profiles in both panels, excepting in the inner most disk where icy solid particles sublimate to vapor. In this section, and in the subsequent Section \ref{sec:disc}, we refer to particles $\ge$ 1 mm in size as pebbles, and $<$ mm as dust. 

\subsection{Varying Gap Location} \label{sec: vary_gap_loc}

For each particle size (described in the next section), we explore gap locations at 7, 15, 30 and 60 au. These radii were chosen to span the outer disk region where ALMA has observed gaps and rings \citep[e.g.,][]{huang18a,long18}, but also are not too far out ($>$ 100 au) that they are unable to trap enough solids to produce a significant difference in vapor abundance in the inner disk. They are also chosen to lie within the disk characteristic radius of R$_{\rm char}$ = 70 au. 
\begin{figure*}[htb!]
	\centering
	\includegraphics[scale=0.45]{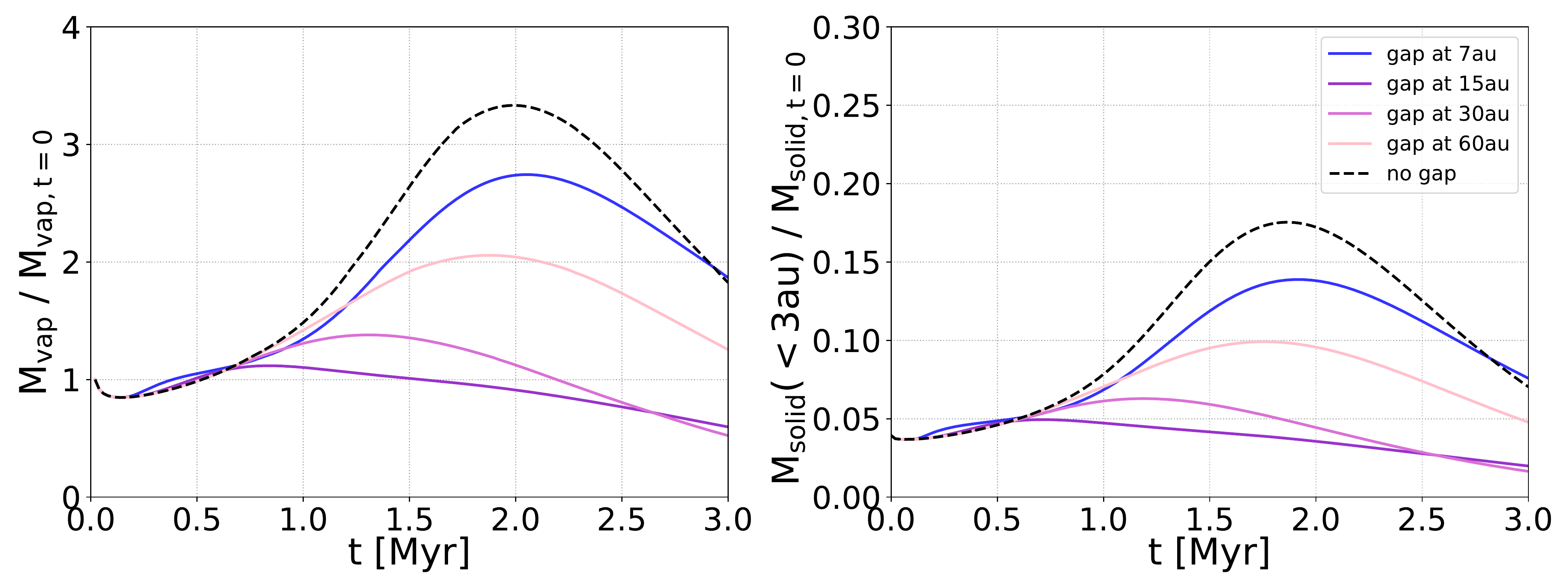}
	\caption{Left panel shows time evolution of water vapor enrichment, i.e., mass of vapor within the snow line when normalized to the initial value at time t=0. Right panel shows time evolution of the fraction of particles within 3 au with respect to total mass of particles in the disk at time t = 0, for our fiducial disk mass (5\% M$_{\odot}$) and fiducial particle size of 1 mm. Different colors show disk models with different gap locations; dashed black line shows the no-gap model. Note here that the rise in each curve is given by icy solid particles that drift in within the snow line and sublimate, thus enriching the inner disk in water vapor; the decline later is instead due to stellar accretion taking over when particles are effectively trapped beyond a gap in the outer disk, or simply drained as in the no-gap case. Note also that the peak enrichment in water vapor is reached at $\sim$ 2.0 Myr for simulations with 1 mm pebbles.}
	\label{fig:vappeb1mm}
\end{figure*}

Figure \ref{fig:vappeb1mm} shows the results of these simulations for the fiducial size of 1 mm (and fiducial disk mass 5\% M$_{\odot}$), where mass of water vapor at time $t$ with respect to initial vapor mass at time t = 0 is plotted in the left panel of Figure \ref{fig:vappeb1mm}, while mass of pebbles delivered within 3 au with respect to total initial pebble mass is plotted in the right panel. For all cases, vapor enrichment (from initial) is at first below 1.0 (when accretion onto the star is dominant) and begins to climb at $\sim$ 0.5 Myr once icy pebbles start to bring water-ice with them from the outer regions, eventually sublimating to vapor within the inner disk. Vapor enrichment reaches a peak at an average time of $\sim2$~Myr for all cases, and then declines with time. This happens because as pebbles are drained from the outer disk, there is no re-supply of ice crossing the snow line, and vapor already present is being accreted onto the star.

For the disk with no gap, vapor enrichment from initial time is the highest as there are no gaps that block delivery of ice-bearing pebbles. All cases with gaps reach relatively lower peak enrichment values. How much these peaks deviate from the no-gap case depends on gap location as follows. For a pebble population all of 1 mm size, water vapor enrichment is minimum when gaps are at 15 and 30 au. On the other hand, the water enrichment that most closely approaches the no-gap model is given by the 7 au (closest) and 60 au (farthest) gaps, though for different reasons. Gaps closer than 15 au are unable to filter 1 mm pebbles and are `leaky', while a gap farther out at 60 au, even if it is an efficient barrier, can trap only a small fraction of the total disk water ice budget, therefore not producing as large an effect in the inner disk water enrichment. 

We also note that the most effective gaps appear to attain their `peak' enrichment earlier in time, than the less effective gaps (i.e., approximately 2.2, 1.9, 1.5 and 0.7 Myr for gaps at 7, 60, 30 and 15 au, respectively) with one notable exception of the 15 au gap. The 15 au gap is the most effective at trapping water but also is slightly leaky, allowing water in at a later time, leading to yet another but slight increase in enrichment. This effect is more prominent in the 3 mm case discussed in the following subsection. 

In the left panel of Figure \ref{fig:vappeb1mm}, we plot the time evolution of incoming pebble mass to total initial pebble mass over time. Similar profiles between the two panels show the connection between the time evolution of the incoming pebble mass and vapor mass, as the vapor mass evolution appears to assume whatever is the shape of the incoming pebble distribution with time. The shape of the incoming pebble mass distribution, i.e., the increase, peak and subsequent decrease in incoming pebble mass over time is a result of two physical processes in action, i.e., the drift of particles as well as the continuous radial diffusion of these particles. This later process seeks to dilute any sharp radial gradients in the radial distribution of particles as they drift inwards from the outer disk. Both of these processes determine the width of the pebble enrichment peak into the inner disk over time. As particle diffusion is dependent on particle size (via the Stokes number), we will see the effect of varying particle size on this peak width in the following subsection. Note that the highest fraction of pebble mass with respect to initial total mass is attained at times slightly earlier than those for vapor, as there is a slight lag for pebbles from 3 au to travel towards the snow line at $\sim$ 0.3 au where they sublimate to vapor.

\begin{figure*}[htb!]
	\centering
	\includegraphics[scale=0.75]{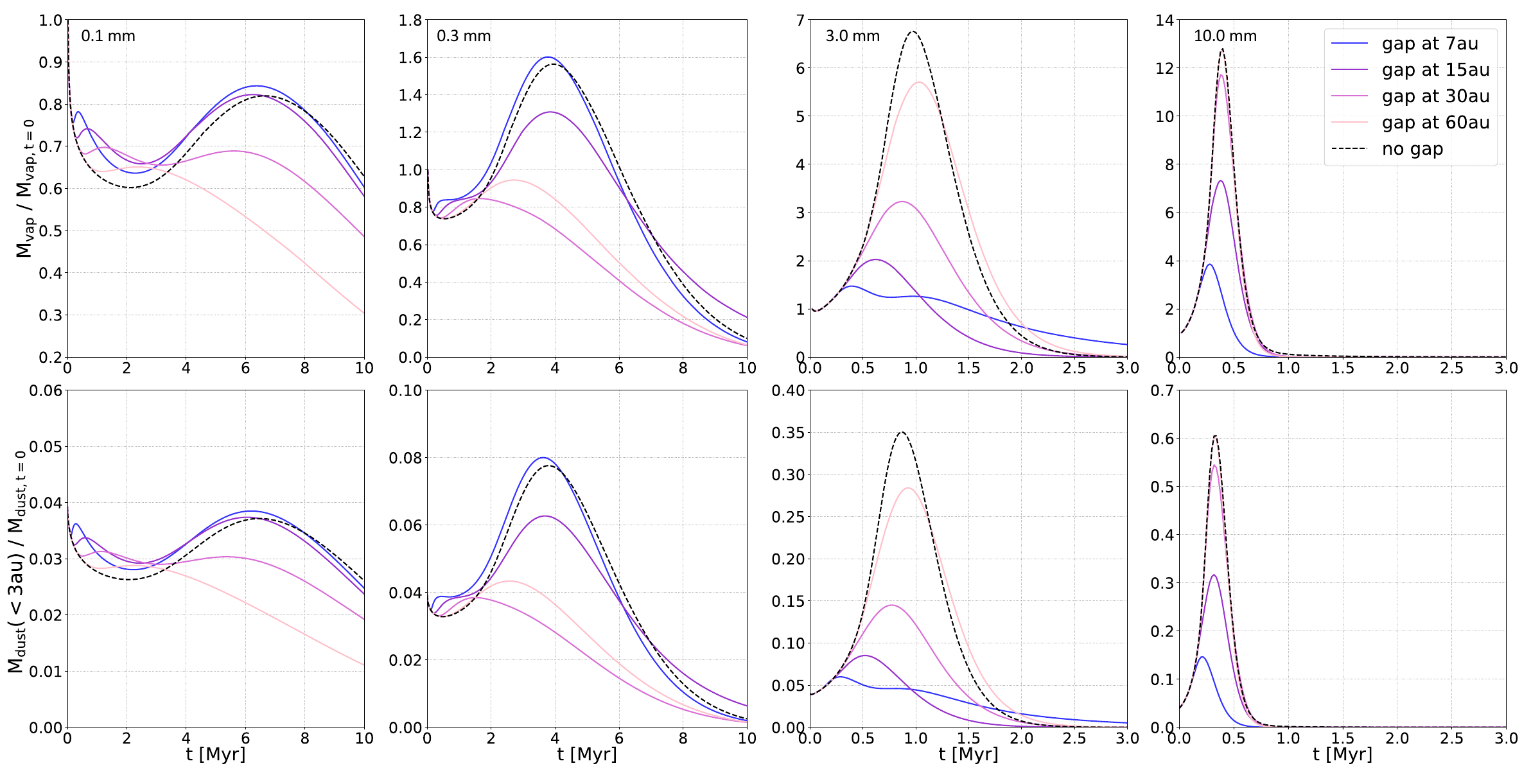}
	\caption{Time evolution of water vapor enrichment (top row) and mass of solid particles delivered within the 3 au inner disk region (bottom row). Plots are similar to those in Figure \ref{fig:vappeb1mm}, here showing models with particle sizes smaller and larger than the fiducial pebble size of 1 mm (with fiducial disk mass 5\% M$_{\odot}$). Line colors denote models with different locations of a gap, varying from 7 au to 60 au, for each particle size. 
	}
	\label{fig:vappebori}
\end{figure*}

\subsection{Varying the Particle Size}
We perform the same suite of runs detailed in Section 3.2 for other particle sizes:  0.1 and 0.3 mm dust particles as well as larger 3 and 10 mm pebbles, results of  which are shown in Figure \ref{fig:vappebori}. For simulations with dust particles, we perform longer simulations (up to 10 Myr), as they take longer to drift in compared to pebbles. We find that for 0.1 mm sized dust particles, 
vapor mass never rises above its initial value at time t=0 throughout the simulation. Vapor mass begins to rise at around 2 Myr as particles carrying ice slowly drift in, raising M$_{\rm vap}$/M$_{\rm vap,0}$ to a peak of 0.8 at only $\sim$ 7 Myr. Simulations with 0.3 mm dust particles see a rise in water enrichment above 1.0 at around 2 Myr, eventually reaching M$_{\rm vap}$/M$_{\rm vap,0}$ = 1.6 at 4 Myr for the no-gap case. Times taken to attain the above peak values in enrichment are much longer for the smaller dust particles cases compared to the fiducial case of 1 mm pebbles as their drift is slower. On the contrary, for pebble sizes $>$ 1 mm the drift is much more rapid, and vapor is enriched significantly from its initial value, reaching up to $\sim$7 and $\sim$13 for the 3 and 10 mm cases with no gap respectively. Also, the peaks of enrichment are reached earlier in time, around 1 Myr for the 3 mm simulations, and $\sim$ 0.4 Myr for the 10 mm simulations.

It is interesting to note which gaps become most effective at trapping water mass as particle size is increased. Assuming that the strongest gap is one that yields minimum vapor enrichment in the inner disk compared to the no-gap case that shows the maximum enrichment, we find that for 0.1 mm particles, gap at 60 au is most effective at blocking most water, and 
for 0.3 mm particles, gap at 30 au is most effective followed by 60 au.
For smaller particle sizes, with the exception of the 60 au gap, all gaps are leaky to varying degrees (see Appendix \ref{app:gapfilter}: Figure \ref{fig:leakage_ori} that shows how effective gaps at different locations are at blocking passage of solids). However, drift is slow enough for 0.1 mm dust particles that a gap at 60 au is able to trap on average $\sim$ 35 \% of the total pebble mass over 10 Myr. Drift is more rapid for 0.3 mm dust particles that gap at 30 au (even though slightly leakier; see Appendix \ref{app:gapfilter}: Figure \ref{fig:leakage_ori}) is able to trap most water mass beyond it over time. Note that for both of these particle sizes, the gap at 7 au produces an enrichment slightly greater than that of the no-gap case. Apart from the `leaky' 7 au gap being unable to filter these dust sized particles effectively, these higher than no-gap enrichments (as well as the small peaks seen around 0.5 Myr for these dust simulations) are due to formation of the gap itself that pushes particles which would have otherwise taken longer (being smaller in size) to drift inward. For simulations with pebble sizes 3 mm and 10 mm, the most effective gaps for blocking water delivery are closer in, compared to the fiducial size of 1 mm. In both cases, gap at 7 au, followed by 15 au, is the most effective. This happens because the 7 au gap is not as leaky for the larger pebble sizes, and traps the most water mass beyond it, therefore producing the minimum water enrichment (i.e. maximum ice-blocking effect). 

Overall, as particle size increases, we see the width of the peak enrichment in vapor (and solid particles) over time become smaller, being broadest for the smallest dust particles (0.1 and 0.3 mm in size), and narrowest in time for simulations performed with 3 and 10 mm sized pebbles. Smaller particles have lower Stokes number (Equation \ref{eqn:diff}) and diffuse more easily within the bulk gas, resulting in broader peak distribution with time, while larger particles do not diffuse as much. The vapor distribution reflects the incoming pebble distribution (shaped by both diffusion and drift) over time.

\begin{figure*}[htb!]
	\centering
	\includegraphics[scale=0.7]{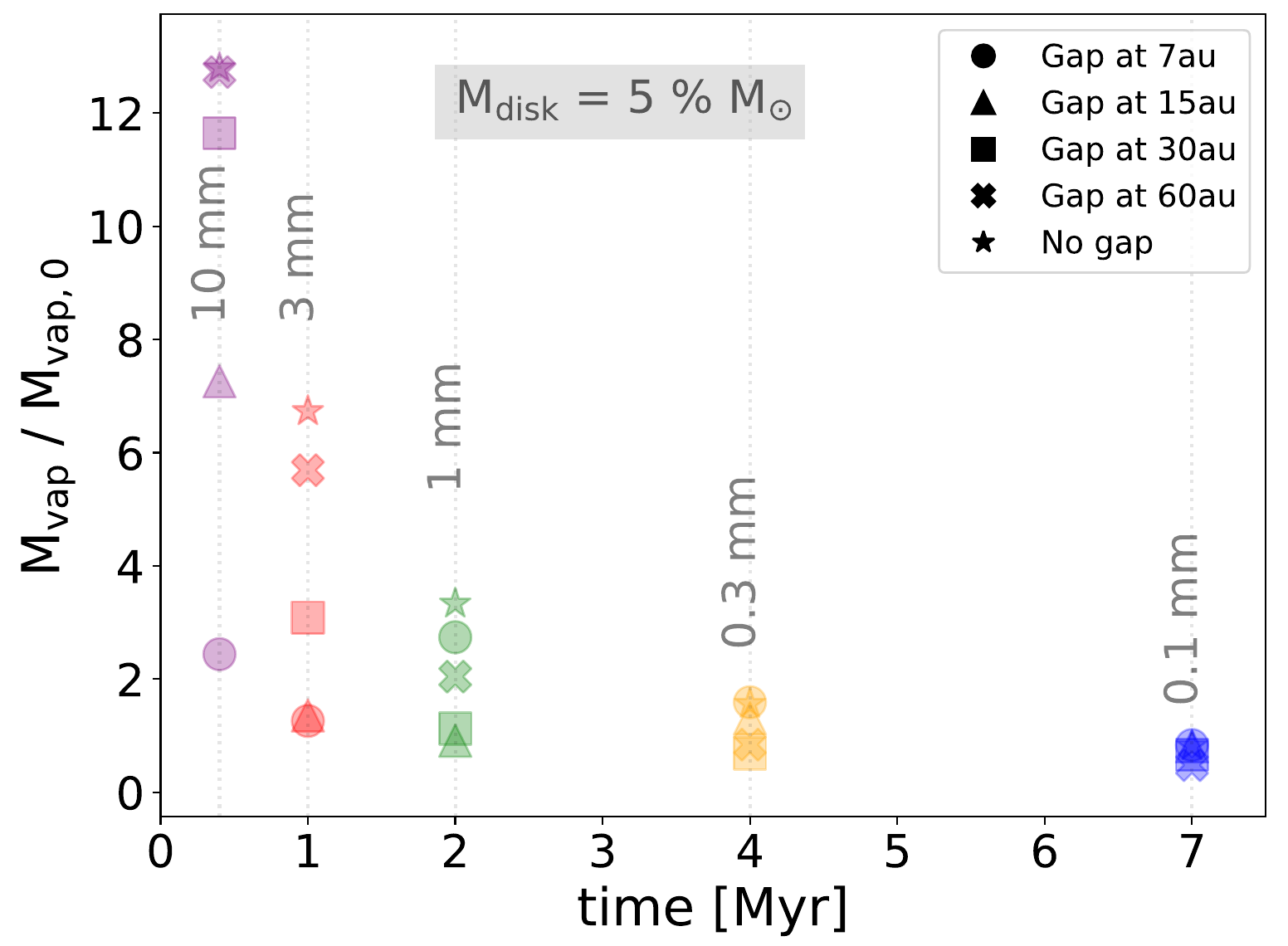}
	\caption{Water vapor enrichment from initial t=0, plotted for each solid particle size, at a specific `peak time' by which most particles have arrived inward of the snow line and have sublimated to enrich the inner disk in vapor. This `peak time' is for 1 cm-sized particles at 0.4 Myr, 3 mm at 1.0 Myr, 1 mm at 2 Myr, 0.3 mm at 4 Myr, and 0.1 mm at 7 Myr. All simulations shown here are performed with the fiducial disk mass (0.05 M$_{\odot}$). Different colors represent simulations performed for each particle size as indicated by labels; different symbols denote particular gap locations.}
	\label{fig:enh_ori}
\end{figure*}

For easier comparison, all of the above results are replotted in Figure \ref{fig:enh_ori} for all 25 simulations discussed so far. For each set of simulations with different particle sizes, we plot vapor enrichment (from initial, i.e.,  M$_{\rm vap}$/M$_{\rm vap,0}$) seen at assumed average time of peak enrichment for each pebble size ( i.e., 0.4 Myr for 10 mm, 1.0 Myr for 3 mm, 2.0 Myr for 1 mm, 4 Myr for 0.3 and 7 Myr for 0.1 mm).

\subsection{Varying Disk Mass}
\begin{figure*}[htb!]
	\centering
	\includegraphics[scale=0.65]{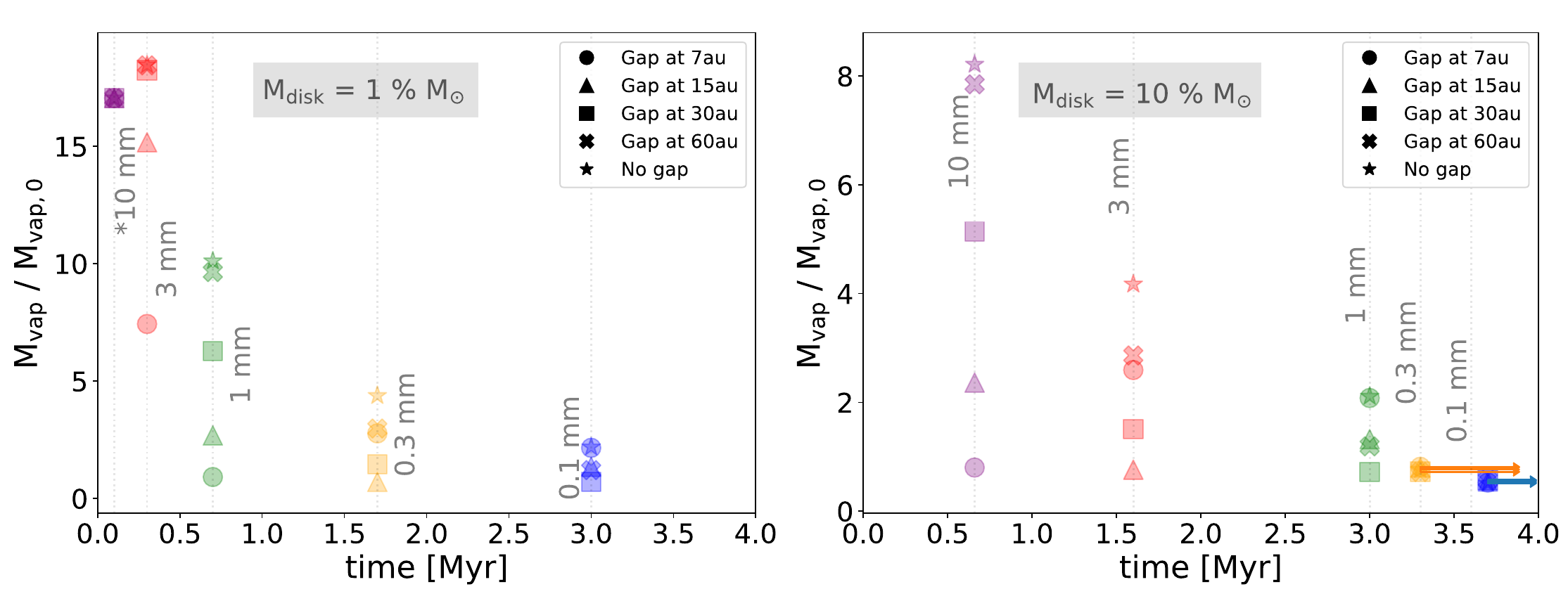}
	\caption{Water vapor enrichment from initial t=0 for simulations of all particle sizes, plotted for disk mass 1\% M$_{\odot}$ (left panel) and 10\% M$_{\odot}$ (right panel). (i) Left panel: solids drift in more rapidly in a less massive disk. All `peak times' are therefore shifted earlier in time. Note here that all 10 mm runs achieve their peak at 0.1 Myr, at the time of gap opening, and therefore show no difference between runs. (ii) Right panel: solids drift in less rapidly in a more massive disk. All `peak times' are therefore shifted later in time.  Note that 0.1 and 0.3 mm dust particles reach their `peak' time at $>$ 4 Myr. Plot symbols and labels are similar to Figure \ref{fig:enh_ori}.}
	\label{fig:enh_1_10}
\end{figure*}

We repeated the 25-run suite of simulations performed in Section 3.3, for the fiducial disk mass, for two additional initial disk masses of 1\% and 10\% M$_{\odot}$, to understand how different disk masses may change the vapor enrichment values obtained so far. The results of these simulations are presented in Figure \ref{fig:enh_1_10}. From these simulations, we find that for less massive disks, solids drift more rapidly inward, and are generally better at trapping solids beyond their gaps compared to simulations with fiducial disk mass.  This results in much higher enrichment for same gap and particle size otherwise, as well as larger deviation in enrichment between gap and no-gap cases. For more massive disks, this trend is opposite. Solids drift inwards more slowly, and gaps being less effective at blocking pebbles produce not only lower enrichments in vapor in the inner disk over all, but also lower deviation in enrichments between gap and no-gap cases for each pebble size.

\subsection{Varying Turbulent $\alpha$}
\begin{figure*}[htb!]
	\centering
	\includegraphics[scale=0.60]{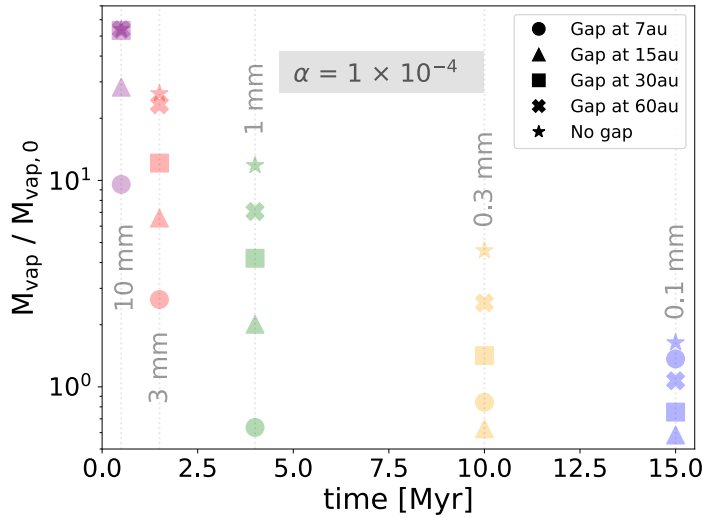}
	\caption{Water vapor enrichment from initial t=0 for simulations of all particle sizes, plotted for the case with $\alpha$ = 1 $\times$ 10$^{-4}$. Note that peak times are shifted to much later times due to slow disk evolution. Vapor enrichments are much higher due to slower diffusion of vapor. Plot symbols and labels are similar to Figure \ref{fig:enh_ori}. Y-axis is plotted in the log scale to capture the wider range of enrichment across particle sizes. }
	\label{fig:enh_alpha}
\end{figure*}

While most disk models typically use a canonical value of $\alpha$ = 10$^{-3}$, observations also suggest that $\alpha$ may be lower in disks, $\sim$ 10$^{-4}$ \citep[][]{pinte16,2018ApJ...869L..46D,huang18a,flaherty20}. We therefore repeated the set of 25 simulations with fiducial disk mass and various particle sizes with lower turbulent $\alpha$ = 1 $\times$ 10$^{-4}$; these results are presented in Figure \ref{fig:enh_alpha}. For simplicity, in all simulations presented in this work, we assume that the same $\alpha$ value regulates both disk evolution as well as the rates of diffusion for particles and vapor  \citep[although it is possible that the $\alpha$ governing each of these processes may be different as explored in][]{pinilla21}. 

While varying $\alpha$ has the overall effect of changing the global rate of disk evolution, it can additionally affect the rates of particle drift, and particle and vapor diffusion as follows. In the presence of accretional heating and the accompanying thermal effect of $\alpha$, lowering $\alpha$ can slow down particle drift via the sound speed c$_s$ \citep[][see Figure 13, and Section 3.2.5]{KD19}. However, in the absence of accretional heating, as assumed in all the simulations presented in this work, lowering $\alpha$ only slows down disk evolution and has only a negligible effect on particle drift speeds (with respect to the gas). On the other hand, lower $\alpha$ significantly decreases the rate of diffusion of both vapor and particles by decreasing vapor diffusivity ${\cal D}_{\rm vap}$ and particle diffusivity ${\cal D}_p$. Among the two, decreasing ${\cal D}_{\rm vap}$ is more significant, as particle drift within the snow line region is now more rapid than diffusion of vapor out of it, allowing more water vapor to stay for longer in the inner disk by slowing down the rate at which it is lost out of the region. This effect is even more prominent for larger particle sizes that drift more quickly and bring in more water.

Thus, for a given particle size, we find that vapor abundance takes much longer to peak in disks with lower $\alpha$ due to slower disk evolution. However, the resulting lower diffusion rates from lower $\alpha$ yield much higher peak vapor abundance than in disks with higher $\alpha$ ($\sim$ 10$^{-3}$), as vapor is lost slowly both to the star and beyond the snow line via back-diffusion.

\section{Discussion} \label{sec:disc}

In this study, we have performed a suite of simulations to explore how substructures in the outer disk, formed as a result of particle trapping beyond a disk gap, may be linked to the level of water vapor enrichment in the inner disk, to test recent findings of an anti-correlation between inner water vapor and outer dust disk radius \citep[][]{banzatti20}. We have used a disk evolution model that includes transport of solids and gas, exploring different gap locations in the gaseous disk and tracking the drifting solid population. Once solids reach the snow line region, the ice sublimates into vapor in the inner disk. We explored a range of particle sizes, gap locations, and disk masses, as well as a lower value of $\alpha$ (affecting disk viscosity, vapor and particle diffusion) to study their influence on the level of vapor enrichment within the inner disk, and determine which parameters produce the minimum enrichment to explain the lower IR water luminosity observed in large disks with substructures \citep[][]{banzatti20}. In this section, we discuss a number of insights gained from the results described in Section \ref{sec:results}.

\subsection{Sweet Spot in Gap Location for Blocking Water Delivery}
A main result from this study is the presence of a `sweet spot' in gap location in the disk for strongly reducing the delivery of icy solids into the inner disk. The sweet spot is due to the combination of two effects. On one hand, the fraction of solids that can be retained in the outer disk decreases with increasing distance of the gap radial location, which leads to less deviation in vapor enrichment in the inner disk, compared to a disk with no gap. On the other hand, a gap located too close-in is `leakier' and not as effective in blocking solid particles drifting from the outer disk (Section \ref{sec: vary_gap_loc}). While most of the solid mass is outside the closest gaps, which could therefore in principle trap the largest amount of icy pebbles, these gaps can effectively reduce the inner water vapor only if they are effective at blocking the solids that arrive at the gap from the outside. This `leakiness' scales with the Stokes number St (Equation 6) and is therefore strongly dependent on particle size and gap location \citep[see e.g., Eq.10 and 11 in][]{pinilla2012}. As seen in Equation 6, St $\propto$ a$_p$ (1/$\Sigma$) within the Epstein limit and St $\propto$ a$_p^2$ r$^{(5/4)}$/$\Sigma$ in the Stokes I limit (with the assumptions for temperature profile used in this work). Thus, in both regimes (assuming a radially constant particle size), gaps are strongest and most effective further away from the star where $\Sigma$ is lower; but these strongest gaps get closer in for larger particle sizes, an effect that is relatively more dramatic in the Stokes regime (with the dependence on a$_p^2$) versus the Epstein regime (with an a$_p$ dependence). 

Overall, these two counter-acting effects of 1) ice mass available at disk radii larger than the gap location, and 2) the gap leakiness, result in the presence of a sweet spot in gap location for blocking the mass of water (in solids) delivered to the inner disk. As gap leakiness is a function of particle size, this sweet spot moves inward as particle size increases. 

Of the four gap locations we use in our simulations, 15 and 30 au located in the middle disk regions provide the best location to see this `sweet spot' effect in operation in our simulations. For simulations with fiducial disk mass = 5\% M$_{\odot}$, we find that 0.3 mm particles show this sweet spot at 30 au, as the closer gaps are too leaky, while the gap at 60 au is not able to trap enough particles beyond them before they drift inward. 1 mm particles have this sweet spot at 15 au (and 30 au), for the same reasons. As this `sweet spot' location is dependent on particle size, particles smaller than 0.3 mm likely have their sweet spot beyond 30 au, while pebbles larger than 1 mm have their sweet spot $<$ 15 au, i.e., 7 au in our simulations. For a smaller disk mass of 1\% M$_{\odot}$, most 0.1 and 0.3 mm dust particles are efficiently blocked by gaps at 15 and 30 au; 1, 3 and 10 mm pebbles are better blocked by the closer-in gap at 7 au. For a larger disk mass of 10\% M$_{\odot}$, the gaps at 15 and 30 au gaps most efficiently block 1 mm and 3 mm pebbles, while larger pebbles (10 mm) will likely be effectively blocked by the 7 au gap. 

For simulations performed with lower $\alpha$-viscosity, we find that generally the above conclusions hold true, deviating only in the fact that a lower $\alpha$ has the effect of pushing the `sweet spot' seen at 15 and 30 au to even smaller particle sizes of 0.1 mm, i.e., even 1 mm and 0.3 mm particles are now well-trapped at the inner gaps of 7 and 15 au.

\begin{figure*}[htb!]
	\centering
	\includegraphics[scale=0.7]{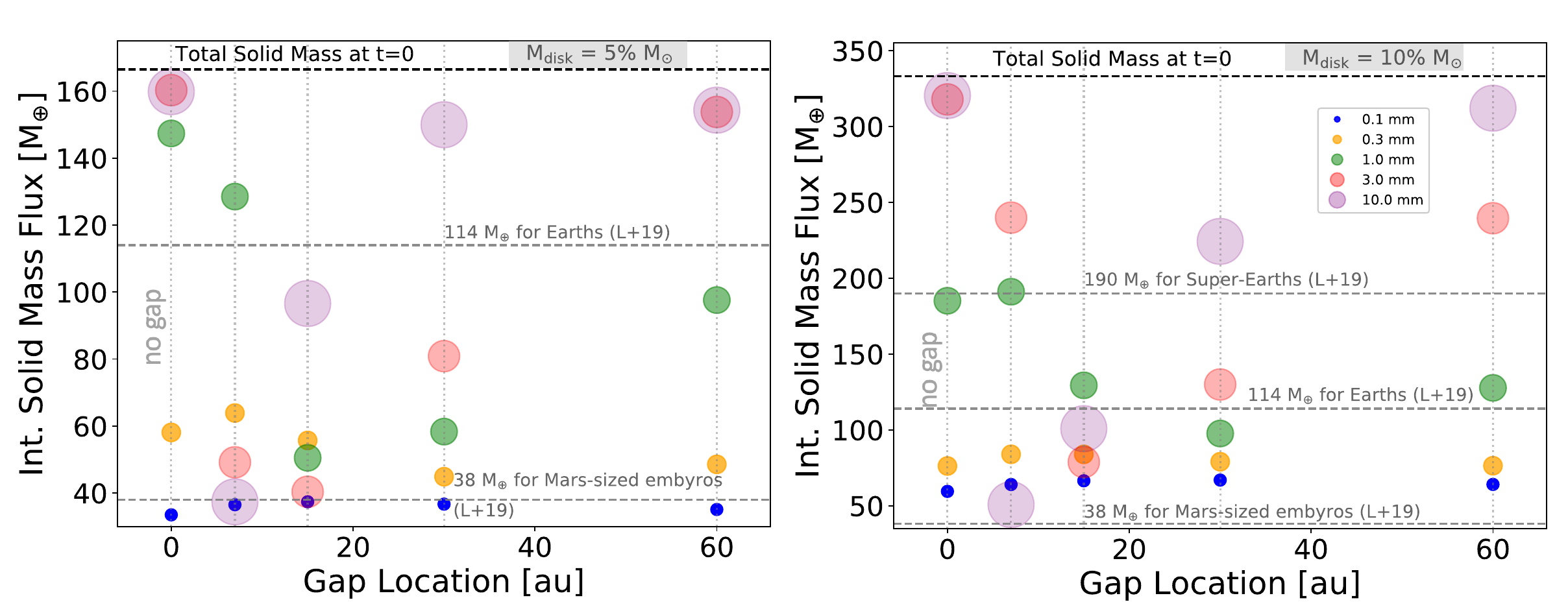}
	\caption{Time-integrated solid mass flux for all 25 runs with different particle sizes, against gap location, for fiducial disk mass 5 \% M$_{\odot}$ in the left panel, and for higher disk mass 10 \% M$_{\odot}$ in the right panel, seen in Figures \ref{fig:enh_ori} and \ref{fig:enh_1_10}. For each simulation, pebble mass flux entering the inner disk region within 3 au is integrated over 3 Myr. Total pebble flux at the start of the simulation including icy and rocky solids for each panel is shown by the horizontal dashed line at 0.01 M$_{\rm disk}$ $\approx$ 166.5 M$_{\oplus}$ for simulations with fiducial disk mass on the left panel, and at $\sim$ 333 M$_{\oplus}$ with higher disk mass on the right panel. Colors and sizes both represent particle size. Following the same sized and colored circle horizontally across gap location also shows the location of the sweet spot for each particle size.}
	\label{fig:intpebmass}
\end{figure*}

Figures \ref{fig:enh_ori}, \ref{fig:enh_1_10}, \ref{fig:enh_alpha} as well as \ref{fig:intpebmass} (which plots integrated pebble mass reaching the inner disk over 3 Myr, and is discussed in detail in Section \ref{sec:planform}) all show this sweet spot effect. Overall, the closer-in gaps effectively block the largest particles that carry the most mass. We also find that as disk mass increases, size of particles effectively trapped by the closer-in gaps (7 and 15 au) increases, i.e., 7--15 au gaps trap particles of size 0.3--10 mm in the 1 \% M$_{\odot}$ disk mass, 1--10 mm pebbles for the 5 \% M$_{\odot}$ disk mass, and 3--10 mm for the 10 \% M$_{\odot}$ disk mass.  

\subsection{Inner gaps are most efficient in blocking most ice-mass}

As explained in the previous section, we find that gaps at 7--15~au are most effective at blocking the largest particles with the most mass from reaching the inner disk; this holds true for a range of disk masses and $\alpha$-viscosity values. These results are consistent with \citet{pin14} that found a correlation between disk cavity radius and spectral index $\alpha_{\rm{mm}}$, suggesting that larger cavity radius (i.e., a more farther-out pressure bump) trapped smaller particles in the bump, while a smaller cavity radius trapped larger particles. These results are also consistent with earlier theoretical predictions from dust-evolution models in \citet{pinilla2012} that found that the critical size of particles that are efficiently trapped at a pressure bump at a disk radius $R$ decreases with $R$. 

If it is the larger pebbles that carry more water ice in terms of mass, even if the smaller dust particles get through, a closer-in gap will be more effective in preventing inner water vapor enrichment. Trapping water ice beyond traps also influences planet formation in the inner disk, which is dependent on the delivery of pebbles. In the following section, we discuss the impact of the delivery of pebbles on planet formation.

We note that this finding is valid for gaps with the properties described in Section \ref{sec:methods} and Appendix \ref{app:gapdepth}; the retention of pebble mass outside the snowline, and in turn its effect in reducing water vapor enrichment in the inner disk, is reduced if gaps are shallower and leakier than what is modeled here. It is therefore very important that future high-resolution  observations reveal the presence and properties of any inner gaps existing in disks at or inside $\sim 10$~au with ALMA or ngVLA \citep[as found in a few disks only so far, e.g.,][]{huang18a,ricci18,andrews16}, to determine how efficient they are in reducing pebble drift and inner water enrichment.

\subsection{Inner Disk Pebble Mass Flux and Planet Formation}\label{sec:planform}

The mass flux of pebbles delivered into the inner disk is critical for planet formation in the inner $\sim$ 3 au disk region. \citet{lamb19} perform numerical simulations to model the growth of planetary embryos by pebble accretion from a range of pebble-mass fluxes, followed by growth of planets from collisions and mergers between these embryos. They explore a range of pebble mass fluxes and find that the growth of Earths and super-Earths in the inner disk is highly sensitive to the pebble mass flux into the inner disk. Specifically, they find that an integrated pebble-mass flux of $\sim$ 114 M$_{\oplus}$ over the first 3 Myr can eventually lead to the formation of Earths after 100 Myr, while $\sim$ 2 $\times$ this mass flux may otherwise eventually lead to the formation of super-Earths.

\begin{figure*}[htb!]
	\centering
	\includegraphics[scale=0.85]{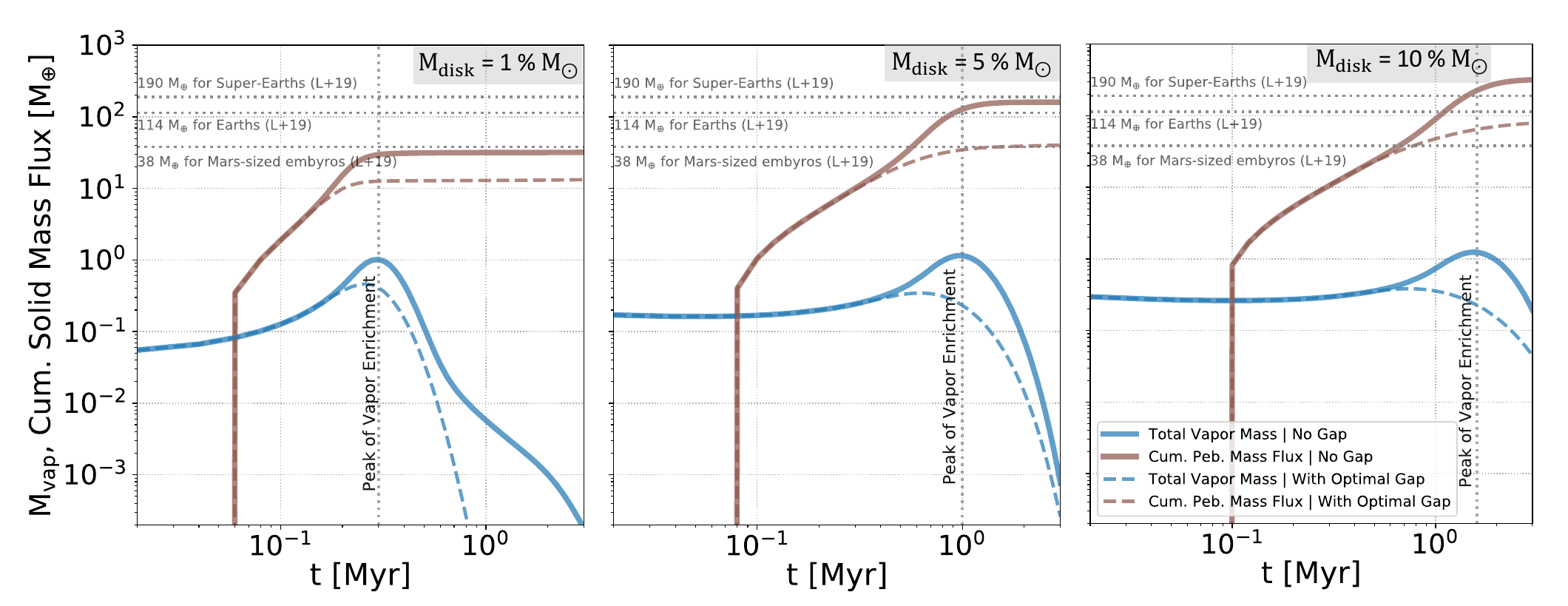}
	\caption{Time evolution of the total vapor mass (in Earth masses; shown in blue) within the snow line region and the cumulative solid mass flux (in Earth masses; shown in brown) entering the inner disk for simulations with 3 mm pebbles. Solid lines represent full-disk models, while dashed lines represent disks with efficient gap. Each panel shows simulations corresponding to different disk masses tested (from left to right: 1, 5 and 10\% M$_{\odot}$). Vertical dotted lines indicate where total water vapor peaks for each case, whereas horizontal dotted lines are integrated pebble mass fluxes from \citet{lamb19}.} 
	\label{fig:cpm_vap}
\end{figure*}

In Figure \ref{fig:intpebmass}, we plot integrated solid mass flux  delivered into the inner disk from the outer disk over 3 Myr, for 25 simulations with fiducial disk mass (5\% M$_\odot$) and 25 simulations with larger disk mass (10\% M$_\odot$), against the radial location of the gap used in these simulations.

Overlaid in Figure \ref{fig:intpebmass} are the range of time-integrated pebble mass flux values computed by \citet{lamb19} (by integrating their Equation 6 over 3 Myr) and listed in their Section 2.5, which eventually lead to the formation of either Mars-sized embryos, Earth-sized planets, super-Earths or more massive planets after 100 Myr in their simulations. Our fiducial disk mass simulations (left panel) begin with the initial total pebble reservoir $\sim$ 166.5 M$_{\oplus}$. For these simulations, we find that the full-disk models with large pebble-sized particles (1 mm or larger) as well as the gapped-disk simulations with leaky inefficient gaps are able to bring in pebble masses into the inner disk greater than the $\sim$ 114 M$_{\rm earth}$ required to eventually form Earths in the inner disk. Otherwise, the presence of efficient gaps reduce inward pebble mass flux to the point that only Mars-sized planets may be able to eventually form in the inner disk via pebble accretion.

Simulations with the larger disk mass (10 \% M$_{\odot}$) have twice the initial solid mass to begin with (equivalent to $\sim$ 333 M$_{\oplus}$). In this case, disks without gaps and with inefficient gaps have the mass to form planetesimals that can eventually form super-Earths, while disks with efficient closer-in gaps may be able to eventually form Earth-like planets, and Mars-sized embryos. (Simulations with the lowest disk mass (1 \% M$_{\odot}$) begin with an initial solid reservoir of only $\sim$ 33 M$_{\oplus}$, and may not be able to form even Mars-sized embryos. They are therefore not plotted in Figure \ref{fig:intpebmass}). 


We note that the pebble mass flux used by \citet{lamb19} was assumed to be a free parameter, and therefore was not linked to gas surface density, as it is in our study.  We also note that our assumption of having solids in a single size in each model is a simplified approach, which should be expanded in future work by including a distribution of particle sizes. Also, the value of $\alpha$ will likely affect the rate of pebble accretion at various times depending on whether pebble accretion is operating in the 2D or 3D regime \citep[][]{sato16,booth17}. We do not consider this effect here as it is beyond the scope of this study. Another assumption in all our simulations is that gaps form at 0.1 Myr. If gaps form at later times in a given disk, more pebbles can drift into the inner disk before gap-opening to provide more solid mass to form planets. We explore this assumption of the time of gap-opening in more detail in Appendix \ref{app:gapopening}.

As our motivation is to link the influx of pebbles to the vapor mass in the inner disk, we also plot in Figure \ref{fig:cpm_vap} the time evolution of the \textit{cumulative} solid mass flux into the inner disk (shown in brown) along with \textit{total} vapor mass in the inner disk (shown in blue). The solid lines show simulations performed with 3 mm pebbles without gap, while the dashed lines show the same simulations with efficient gap. Different panels show the range of disk masses (1, 5 and 10 \% solar mass) that we tested in our simulations, where each has different initial total pebble mass reservoirs as mentioned above. 

As we already see in previous plots, we note that the total vapor mass initially increases with time, attains a peak, and then decreases with time as vapor gets accreted onto the star. Cumulative pebble mass profiles show that vapor mass peaks around the same time a large fraction of pebbles have entered the inner disk. If there is an efficient gap present, then both vapor and pebble mass do not attain the high enrichment that would have been possible in a disk without a gap. The presence of an efficient gap can thus hinder the formation of large enough embryos in the inner disk via pebble accretion, thwarting the eventual formation of larger planets from mutual collisions and mergers of these embryos. If all the solid mass were in 3 mm pebbles, then comparing with the integrated fluxes of \citet{lamb19}, the presence of an efficient gap may hinder the formation of Earths in a disk with an initial disk mass of 5\% M${_\odot}$, and of super-Earths in a disk with an initial disk mass of 10\% M${_\odot}$. Moreover, vapor enrichment in the inner disk is highly time-varying and our models suggest that the inner disk becomes water-poor after pebble drift declines, with the remaining water vapor being accreted onto the star (Figure \ref{fig:cpm_vap}). We predict that for sub-Neptunes and larger planets that may form relatively late and inside the snow line will have accreted water-poor, i.e., dry atmospheres in disks with or without a gap in the outer disk outside of the snowline \citep[though slightly more in disks without a gap, see also][]{bitsch21}.

We argue that even if we included a more realistic solid particle population with a range of sizes, rather than just one fixed size across the disk, it is likely that the inner disk gaps still play a key role in hindering the formation of larger planets via pebble accretion, as they block the largest particles with the most mass. This result is consistent with the study of \citet{vdmm21}, where they link known exoplanet demographics with large disk surveys, and conclude that super-Earths (especially those around low-mass stars) must have formed in small disks that have no drift-blocking gaps. Moreover, we also argue that as vapor enrichment in the inner disk is closely related to the incoming pebble population, warm water vapor abundances may not only provide an estimate of the pebble population that has already entered the snow line region, but may also be useful to infer the presence of any unseen gaps close to the water snow line, but too close-in to be resolved with ALMA.

In connecting inner disk pebble mass flux and presence of disk structure, our solar system also provides an important case study. While the solar system currently hosts two Earth-like planets, meteoritic evidence suggests that the solar nebula itself may have had a gap in the protoplanetary disk within 1 Myr of formation (between 0.4 - 0.9 Myr) at around 3 au, due to the formation of Jupiter's core \citep{krui17}. As mentioned before, while its possible that very close-in gaps may be leakier to even pebbles, a later gap-opening would allow for higher pebble mass flux into the inner disk at earlier times, yielding overall higher time-integrated pebble mass fluxes (see Appendix \ref{app:gapopening}). Moreover, even if Jupiter's growth blocked almost all pebble delivery into the inner disk, slower mechanisms of planetary growth such as planetesimal formation may eventually yield Mars-size embryos that can grow to terrestrial planets.

\subsection{Comparison to the Observed trend}

\begin{figure*}[htb!]
	\centering
	\includegraphics[scale=0.75]{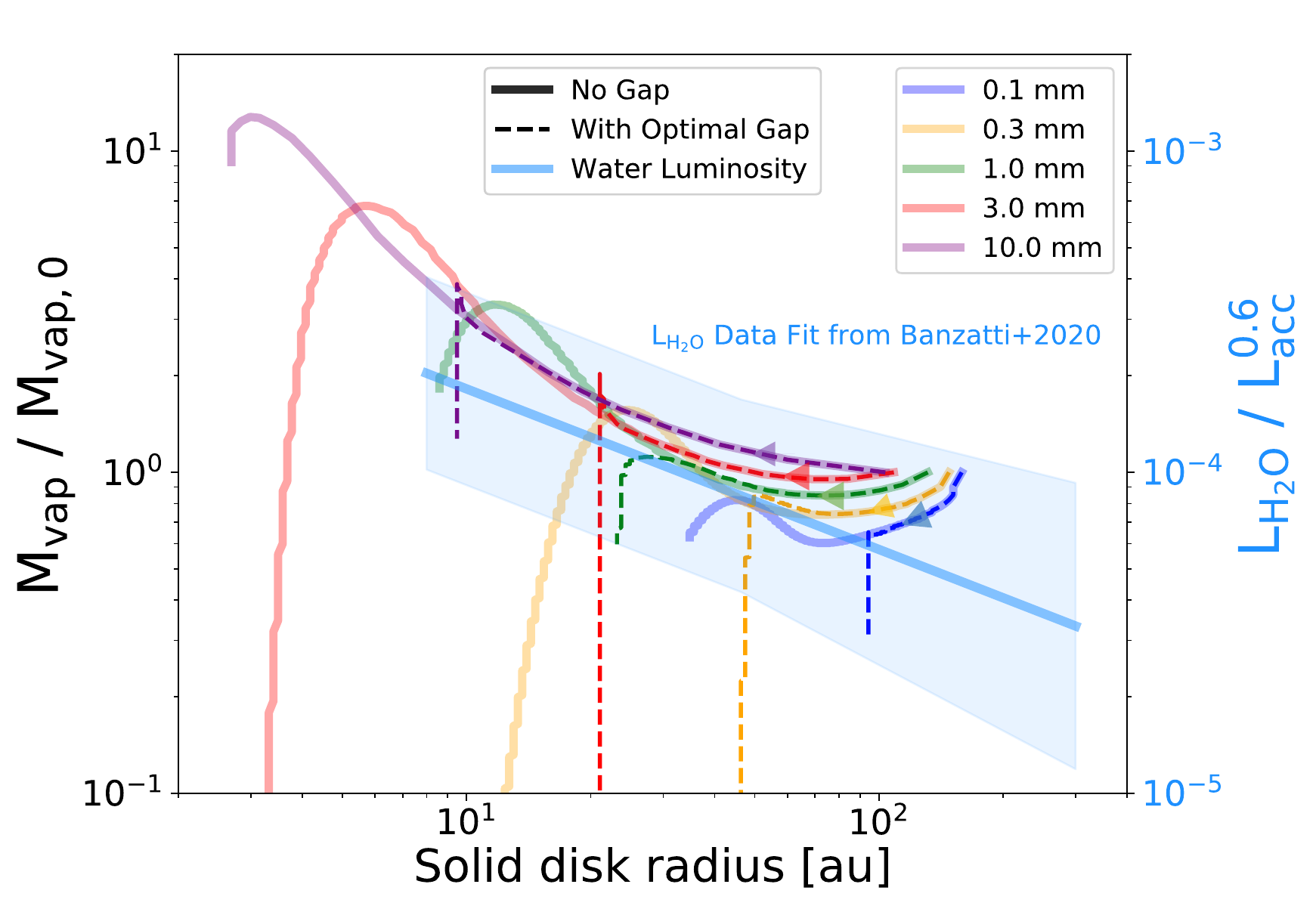}
	\caption{Evolutionary tracks of vapor enrichment (over initial time) plotted against solid disk radius (enclosing 90\% solid mass) for disk models with each particle size, one with no gap, and one with efficient gap that provides the best barrier for water ice delivery. Models with no gap are denoted by solid lines, while models with gap are denoted by dashed lines; arrows indicate direction of progression of time of these evolutionary tracks. Color denotes pebble size used in the simulation, similar to previous plots. Simulations shown here are performed with fiducial disk mass of 5\% M$_{\odot}$. All runs are plotted until end of their simulation times, i.e., 10 Myr for 0.1 and 0.3 mm dust particle simulations., and 3 Myr for 1 and 3 mm pebble simulations, excepting for the case with 10 mm pebble simulations, shown until only 0.5 Myr; (by this time the pebble disk has almost completely depleted). The second y axis in blue shows the linear regression fit (light blue solid line; blue shaded region for the uncertainty) to measurements of water luminosity L$_{\rm H_2O}$ (corrected for accretion luminosity L$_{\rm acc}$) in a sample of disks with disk size estimates from resolved ALMA observations from \citet{banzatti20}. A scaling factor of $10^4$ is applied between the two y axes for comparison purposes only.}
	\label{fig:pebdisk}
\end{figure*}

In Figure \ref{fig:pebdisk}, we plot the vapor enrichment against solid disk radius for each time over 3 Myr, for both a full-disk model and the most effective gap model for each particle size (with fiducial disk mass). Here we compute the solid disk radius to be the effective disk radius that encloses 90 \% of the solid mass, to mimic the observationally-defined disk radii based on 90-95\% of the millimeter flux \citep[e.g.][]{long18,pin20}. All disks start at the same radial size and same inner vapor mass. For each particle size, the evolutionary track for a disk with no gap eventually attains a higher water enrichment with time than a disk with a gap, as shown in the previous sections. Simulations with larger particle sizes also show higher inner water enrichment, due to rapid drift of larger pebbles carrying more ice mass. A disk with a gap has a larger final solid disk radius than one without a gap, for each particle size, which is both consistent with other models \citep{pin20} as well as observations \citep{long18}. Disks become very small ($\sim$ 10 - 20 au within 1 Myr with a gap, or 2-3 au without a gap) if the disks only had rapidly drifting pebbles (as seen in the 3 and 10 mm cases) rather than slowly drifting dust particles which maintain much larger disks (around 30 au without gap and around 100 au with gap) over 10 Myr, as seen for 0.1 mm dust particle simulations.

This overall trend is broadly consistent with the findings in \citet{banzatti20} (see their Figure 6), of an anti-correlation between inner disk water luminosity from infrared spectra and outer pebble disk radius from millimeter interferometry. In Figure \ref{fig:pebdisk}, we overlay the linear regression fit from \citet{banzatti20} for comparison to model vapor enrichment values from our simulations, by applying a scaling factor of 10$^4$ between the two y-axes. This scaling factor is arbitrary here and only applied to compare the slopes of the trends from observations and models; the relation between inner disk water mass enrichment and infrared water luminosity is still currently unknown and should be investigated using thermo-chemical disk models \citep[e.g.][]{Woitke18}.

In their work, \citet{banzatti20} propose that the observed trend in water luminosity could be interpreted in terms of water abundance, with small disks ($<$ 60 au) having a higher water abundance compared to larger disks (60 au $<$ R$_{\rm dust}$ $<$ 300 au). From the results of our simulations, we further argue that this correlation arises because compact disks have experienced more drift and thus increased water abundance in the inner disk. On the contrary, large disks have gaps and rings, as observations show, and these gaps are presumably able to retain their large outer disk pebble reservoirs by trapping solid particles beyond them and halting their rapid drift into the star, resulting in decreased water abundance in the inner disk. Water abundances are still very uncertain as based on the spectrally-blended Spitzer data. Going forward, spectra taken by the James Webb Space Telescope (JWST) will be able to better spectrally resolve individual emission lines with different optical depths \citep[e.g.,][]{notsu17,greenwood19}, and therefore significantly improve measurements of water abundances in inner disks, as well as better resolve the location of the water snow line at the mid-plane. This has been attempted with ALMA but so far without success \citep[][]{notsu19}.


Finally, an important caveat of our present work is that we only include one gap in our disk models, resulting in smaller outer disk radii in comparison to observations of disks where multiple gaps and rings are often found \citep{andrews20}. From our results in this work, we predict that while the outermost gaps are most important for setting the disk pebble radius that is observable with ALMA, the inner most gaps may be the most important for blocking the most water from the inner disk. In a future work, we will study the case of disks with multiple gaps, fully exploring the link between pebble/dust disk size (in disks with multiple gaps) and inner disk water abundances.

\subsection{Comparisons with Other Works}
We note that in this study, we make a number of assumptions regarding the solid population. We considered a 50-50\% ice-to-silicate ratio in the incoming solid population, to emulate solid particles with a rocky inner core and an icy mantle. We also assume only one size of particles is present at each location, and further assume this particle size is radially constant and does not change with time, neglecting any size evolution due to growth or fragmentation, which has been included in works by \citet[e.g.,][]{birnstiel12}, or assuming a single dominant particle size at each radius but otherwise varying over time and radial location \citep{schoor17,schoo18}. Since our pressure bumps are mainly located in the drift-limited regime as in \citet{birnstiel12}, the maximum particle size would also decrease with radius as well as time (as solid-to-gas ratio decreases), resulting in gaps that filter or allow different particle sizes at different times. These assumptions, though significant, are aimed at simplifying our models leading us to unique insights presented in this paper.

However, based on the results of \citet[][]{pin14}, we argue that even if we had considered a full treatment of the dust evolution in the drift limited regime \citep[][]{birnstiel12}, we predict that the main conclusions of this study will still hold, i.e., that for each particle size, there will be some unique regions in the disk that will be more efficient than other locations at blocking water delivery in particles of those sizes into the inner disk, and that overall, the innermost gaps will be best at blocking most water mass into the inner disk.  

Several recent works have investigated the effect of radial drift on other molecular species, in some cases combining dust evolution with ongoing chemical processing \citep[e.g.,][]{booth19,krijt20}. In the case of CO, observed enhancements of gas-phase CO interior to the CO iceline can be compared to similar models to estimate pebble fluxes at the CO iceline (typically at several tens of au) \citep{zhang19,zhang20}. For water in the midplane, the effects of chemical processing appear to be small compared to sublimation/desorption following pebble drift \citep[][Fig.~4]{booth19}. For other species such as CO, both might be important, with chemical processing quickly removing CO from the inner $\sim$ au in the absence of drift of icy particles. We note also that we have not modeled the effects of radial drift and gap opening on the disk's thermal structure \citep[][]{alarcon20,vdm21}, although this could potentially alter the condensation and sublimation behavior - especially for the most volatile species \citep[e.g.][]{Cleeves16}.

Finally, we also make the overarching assumption that trapping of pebbles in pressure bumps beyond gaps is the main mechanism by which pebbles are prevented from drifting inwards into the inner disk. Grain growth and subsequent planetesimal formation may also lock up ices in the outer disk \citep[][]{mcclure19}. We will further explore these effects and their interplay in a future work.

\section{Summary and Conclusions} \label{sec:conc}
In this work, we have studied the link between the dynamics of solid particles drifting inward from the outer disk and the vapor enrichment in the inner disk. We used a volatile-inclusive disk evolution model, which tracks the distribution of water in both ice and vapor throughout the disk. We included structure in the outer disk in the form of a gap, and explored the effect of various parameters such as particle size, gap location, initial disk mass and $\alpha$-viscosity on the time-evolving vapor abundance in the inner disk. We list the main results of this study as follows: 

\begin{enumerate}
    \item In agreement with previous works, the water vapor abundance in the inner disk evolves with time: the inner disk is enriched in vapor from initial time (t=0) when icy solids drift inward from the outer disk and ice sublimates within the snow line. At later times, when the influx of icy solids decreases, water vapor is depleted by accretion onto the star. 
    \item Gaps in the outer disk trap particles in the pressure bump beyond it. For each particle size, there is a sweet spot in gap location for blocking water-ice delivery into the inner disk. If a gap is too far, not enough water is blocked to produce a significant reduction in vapor enrichment in the inner disk; if a gap is too close-in, it is `leakier' (depending on particle size) and therefore not as effective in blocking icy particles of that size beyond the gap.
    \item Inner gaps at $\sim$~7--15 au are the most efficient in blocking most water mass from entering the region of the snow line, as they are able to efficiently block ice-bearing particles as large as 3 mm and 1 cm, which drift faster and carry more ice mass than smaller particles. Gaps at 30 au are more efficient in blocking smaller particles, but these deliver a lower ice mass to the inner disk and not as quickly.
    \item Comparing models with initial disk mass = 5 \% M$_{\odot}$ with the results of \citet{lamb19}, we find that disks without gaps or with leaky inefficient gaps may be able to form exo-Earths; on the contrary, disks with efficient gaps may be unable to form anything larger than Mars-sized planets via pebble accretion. For simulations with larger disk mass =  10 \% M$_{\odot}$ (and thus a larger solid reservoir), disks without gaps or with inefficient gaps may eventually form Earths and super-Earths in their inner disk regions, whereas disks with efficient gaps may only be able to form Earth- or smaller Mars-sized planets.  In general, we predict that disks with inner gaps may not be able to eventually produce Earths or super-Earth planets.
    \item Disks with lower $\alpha$ may show much higher vapor enrichments in the inner disk as compared to disks with higher $\alpha$. For a fixed particle size, decreasing $\alpha$ changes the relative importance of drift over vapor diffusion; particle drift is negligibly affected by lower $\alpha$ while vapor diffusion is strongly decreased. This allows for vapor to persist in the inner disk for longer times.
    \item This exploration shows that multiple factors affect water abundance in inner disks. For disks that have similar initial disk mass and similar $\alpha$, since vapor abundance in the inner disk follows the pebble mass flux entering the inner disk, we argue that water vapor observations could be used as a proxy for estimating pebble mass flux coming into the innermost regions of disks \citep[][]{banzatti20}.

\end{enumerate}

\acknowledgments
We thank Steve Desch for helpful discussions, and anonymous referees for helpful suggestions that improved this manuscript.


\bibliography{sample63}{}
\bibliographystyle{aasjournal}

\appendix

\section{Gap properties across disk radii and Disk Mass}\label{app:gapdepth}

\begin{figure*}[htb!]
	\centering
	\includegraphics[scale=1.20]{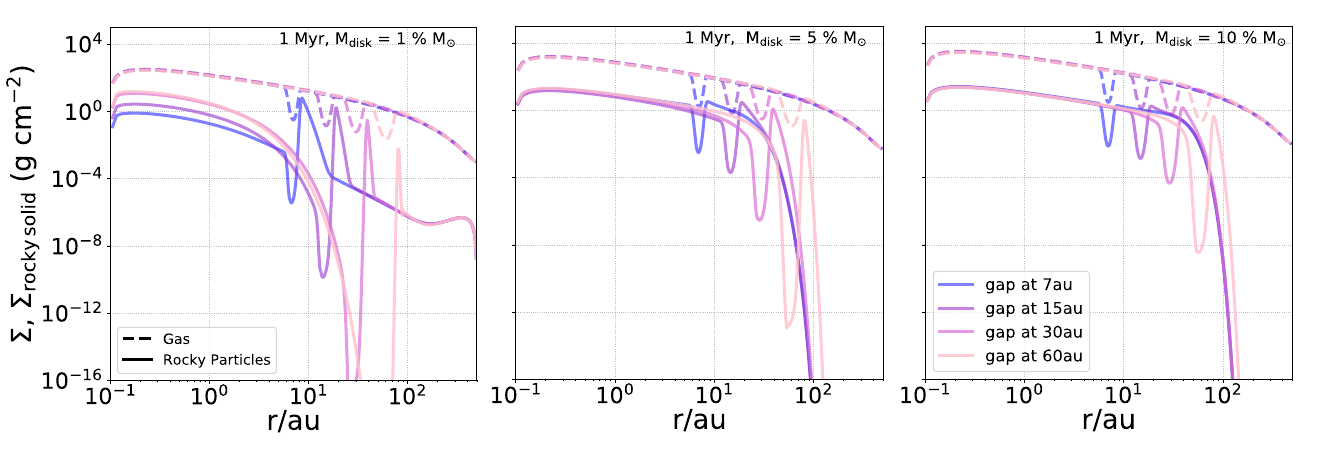}
	\caption{Surface density profiles of gas (dashed) and rocky solid particles (solid lines) across radius $r$ for each disk model with a different gap radius, to highlight depth of gaps at different radii. All curves are shown at 1 Myr, for simulations performed with 1 mm pebbles. Left, middle and right panels show the same simulations performed for disk masses 1 \%, 5 \% and 10 \% M$_{\odot}$. Colors denote disk models with gaps at different radii as shown in previous plots.}
	\label{fig:gap_depth}
\end{figure*}

The properties of gaps in disks are sensitive to multiple aspects of the physical environment of the disk in which they form \citep[e.g.][]{vdm19,zhang18,2018ApJ...869L..46D}. In our models, while we initially impose the same perturbation in the $\alpha$ profile to create a gaseous gap across all our models, the solid particle gap depth and width are dependent on the surface density of gas (and therefore radial location of gap, and the initial disk mass assumed) as well as time (as both surface density of gas and particles evolve with time). In Figure \ref{fig:gap_depth}, we show the radial profiles of surface density in gas and solids (rocky solid particles) for disk models with drifting 1 mm particles at time t = 1 Myr for different disk masses. As shown in Figure \ref{fig:gap_depth}, we find that for the type of gaps defined in Section \ref{sec:methods}, gap depth and width are larger at larger disk radii, and with lower disk mass. We also find that the gap depth increases with time (not shown here).  

\section{How different gap locations filter particles of each size}\label{app:gapfilter}

\begin{figure}[htb!]
	\centering
	\includegraphics[scale=1.0]{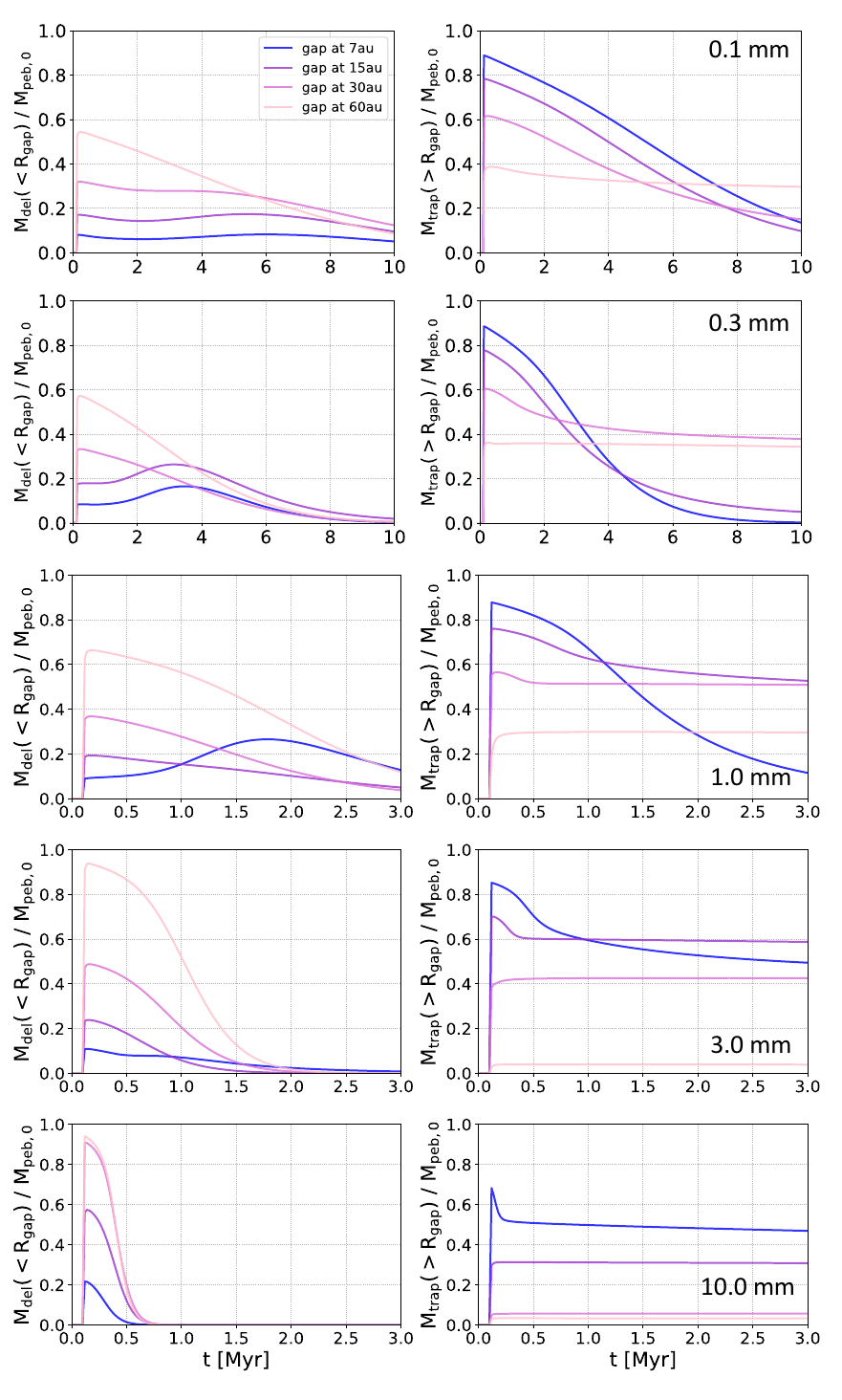}
	\caption{Plots showing mass of pebbles delivered inwards to the inner disk within $R_{\rm gap}$, and mass of pebbles trapped beyond $R_{\rm gap}$, as a fraction of total initial pebble mass, with time, for fiducial disk mass 5\% M$_{\odot}$. Each row shows simulations for each pebble size. Note that we performed longer simulations (10 Myr) for 0.1 and 0.3 mm dust particles.} 
	\label{fig:leakage_ori}
\end{figure}

In Figure \ref{fig:leakage_ori}, we show the mass of solid particles delivered inwards into the inner disk or trapped outside of the gap as fractions of the total initial solid mass. This plot is useful to observe how gaps at different radial locations filter out particles of each size. For example, for the fiducial case shown in Figure \ref{fig:vappeb1mm}, the third row of plots in Figure \ref{fig:leakage_ori} show how the gap at 7 au (shown in blue) is ineffective as a barrier and `leakier' than gaps at other locations. Here, the left panel shows a surge in delivered pebble mass after 1 Myr, not seen for other gaps. Likewise, the right panel shows the trapped mass of pebbles (in the pressure bump beyond the 7 au gap) declining with time.

\section{Effect of Accretional Heating and a farther-out Water Snow Line Location}\label{app:accheat}

\begin{figure*}[htb!]
	\centering
	\includegraphics[scale=0.45]{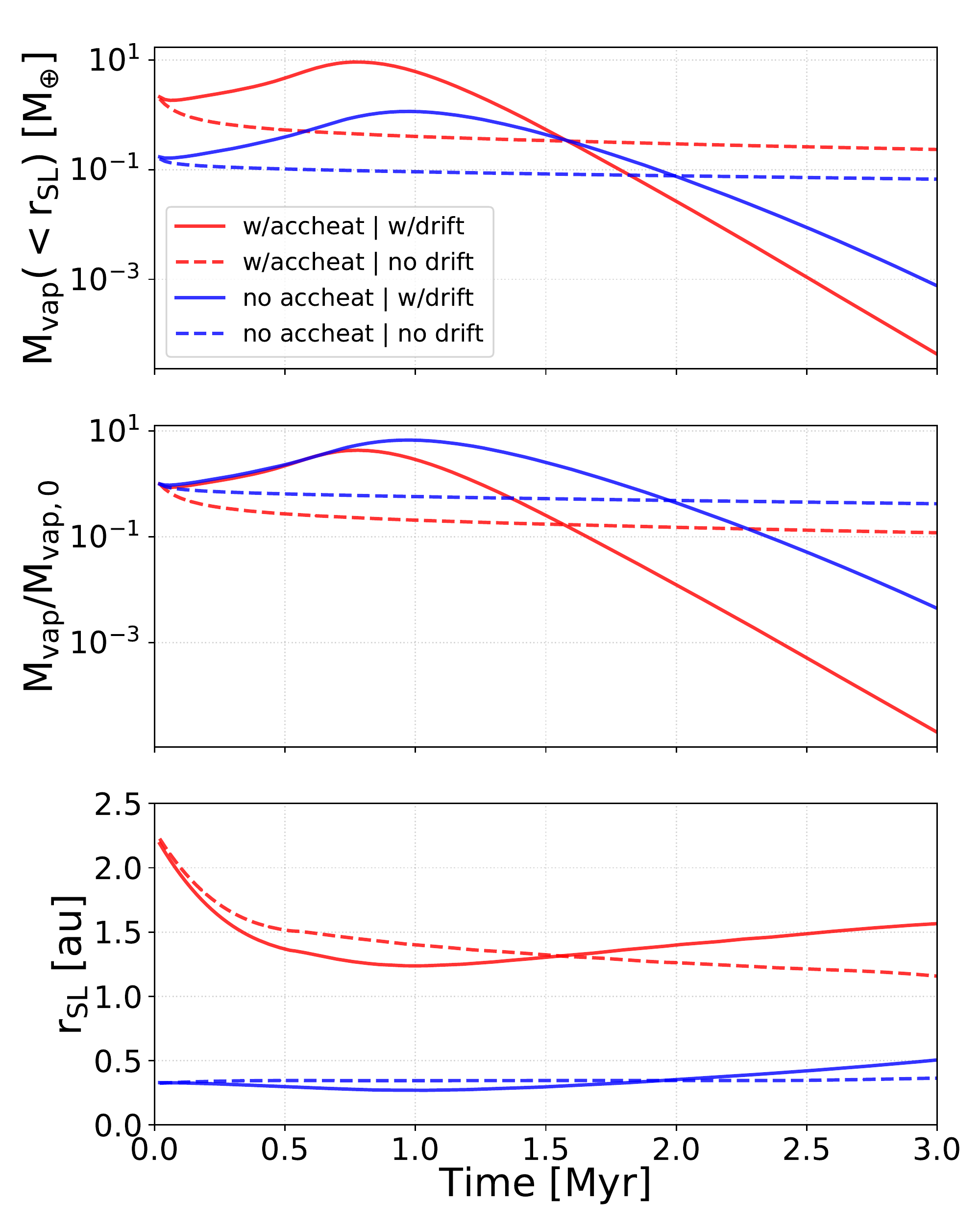}
	\caption{Time evolution of total vapor mass, vapor mass normalized to initial time t=0, and the radial location of the snow line are plotted in top, middle and bottom panels, respectively. Simulations with only passive heating (as described throughout the paper) are shown in blue, while simulations including accretional heating are shown in red. Solid lines represent simulations where drift of particles are allowed, while dashed lines represent those where particle drift is halted (though particle diffusion is still present). All these simulations are performed with 3 mm pebbles and without any gaps in the outer disk. Accretional heating has the effect of not only pushing the snow line outward, but also of evolving the inner disk more rapidly, such that vapor mass is more quickly depleted at later times, compared to simulations with only passive heating. Note that slight changes in r$_{\rm SL}$ between drift and no-drift cases are due to change in the radial vapor pressure gradient with and without drift of icy solids.} 
	\label{fig:accheat_summ}
\end{figure*}

In this study, we neglected the effect of accretional heating on the thermal structure of the disk. As a result our snow line is much further in (at around 0.3 au) and is relatively fixed in radius. This was intended as we wanted to first understand how the mass of vapor changed within a fixed snow line radius. All simulations in our study were therefore performed with a thermal profile that was purely influenced by passive starlight.

We performed a few simulations with accretional heating to understand the impact of this additional heating source on snow line radius r$_{\rm SL}$ and the time evolution of vapor mass in the inner disk. We included accretional heating as follows:

\begin{equation}
{\rm T}_{\rm acc}(r) = \left[ \frac{27}{128} \, \frac{k}{\mu \sigma} \, 
 \Sigma(r)^2 \, \kappa \, \alpha(r) \, \Omega(r) \right]^{1/3},
\end{equation}

where $\Sigma$ is gas surface density, $\alpha$ is the turbulent viscosity, $\Omega$ is Keplerian angular frequency, and we assume a population of fine dust to be our source of opacity yielding $\kappa$ = 5 cm$^2$/g. Mean molecular weight $\mu$ is assumed to be 2.33 $\times$ proton-mass, and $k$ and $\sigma$ are Boltzmann and Stefan-Boltzmann constants respectively. (See \citetalias{KD19} for more details). We add the contributions of T$_{\rm acc}$ and T$_{\rm pass}$ as follows: T(r) = $\left[ {\rm T}_{\rm acc}(r)^4 + {\rm T}_{\rm pass}(r)^4 \right]^{1/4}$, to obtain the total T(r). 

We have performed four simulations with fast-drifting 3 mm particles where we tested the effect of the presence and absence of accretional heating, in disk models where we either allow or do not allow particles to drift inwards, and show our results in Figure \ref{fig:accheat_summ}. The top panel shows the time evolution of total vapor mass within the snow line region, the middle panel shows the time evolution of vapor mass within the snow line region normalized to its initial value (as done in previous plots) and the bottom panel shows the movement of the snow line with time. Simulations with and without accretional heating are shown in red and blue lines respectively. Simulations with drift (solid lines) show the familiar increase to a peak in vapor mass and decrease later, as seen in previous plots. Simulations without drift (dashed lines) show a profile that is relatively constant in time. Note that even though we have removed drift in these cases, particle diffusion is still present that continuously brings in some icy pebbles into the inner disk throughout the simulation, leading to a small yet constant source of vapor mass throughout. (Simulations with drift and diffusion use up the outer icy pebble reservoir more rapidly and efficiently than simulations with only diffusion; this is the reason why simulations without drift have higher vapor enrichment than those with drift at later times.)

Including accretional heating has the following main effects: i) a farther placement of the snow line and therefore higher total vapor mass in the inner disk; and ii) more rapid evolution of the inner disk. As Figure \ref{fig:accheat_summ} shows, a disk model with accretional heating has r$_{\rm SL}$ initially at $\sim$ 2.25 au, which subsequently moves inward to around 1.25 au at around 1 Myr where the vapor mass peaks. A disk with a far-out snow line has much higher vapor mass at any time, than a disk with a snow line closer-in, as evident in the top panel. Normalizing with the initial vapor mass at t=0 allows us to more easily compare between these two cases, and note that while the normalized peak of vapor mass without accretional heating is slightly higher than with accretional heating, but more importantly, both total and normalized vapor mass decline more rapidly with time if accretional heating is included. This is because a disk with accretional heating evolves more rapidly than a disk with only passive starlight and therefore depletes water vapor from the disk more rapidly. In our simulations, we find that while accretion rates changed from 1.2 $\times$ 10$^{-8}$ M$_{\odot}$/yr at t=0 to 1.8 $\times$ 10$^{-9}$ M$_{\odot}$/yr at 3 Myr  for the case with accretional heating, for simulations with passive heating only, accretion rates changed from 6.0 $\times$ 10$^{-9}$ M$_{\odot}$/yr at t=0 to 1.8 $\times$ 10$^{-9}$ M$_{\odot}$/yr at 3 Myr. We conclude that including accretional heating has the effect of increasing vapor mass within the snow line region, which is also depleted somewhat more rapidly with time.

We have also performed simulations where we have tested disk models with only passive heating but with higher overall temperature, i.e., T$_0$ $>$ 118K in Equation 5, such that r$_{\rm SL}$ is at 1 au instead. We find that the main conclusions of our work remain unchanged with the change in the location of the snow line.

\section{Effect of Gap-opening at 1 Myr}\label{app:gapopening}

\begin{figure}[htb!]
	\centering
	\includegraphics[scale=0.4]{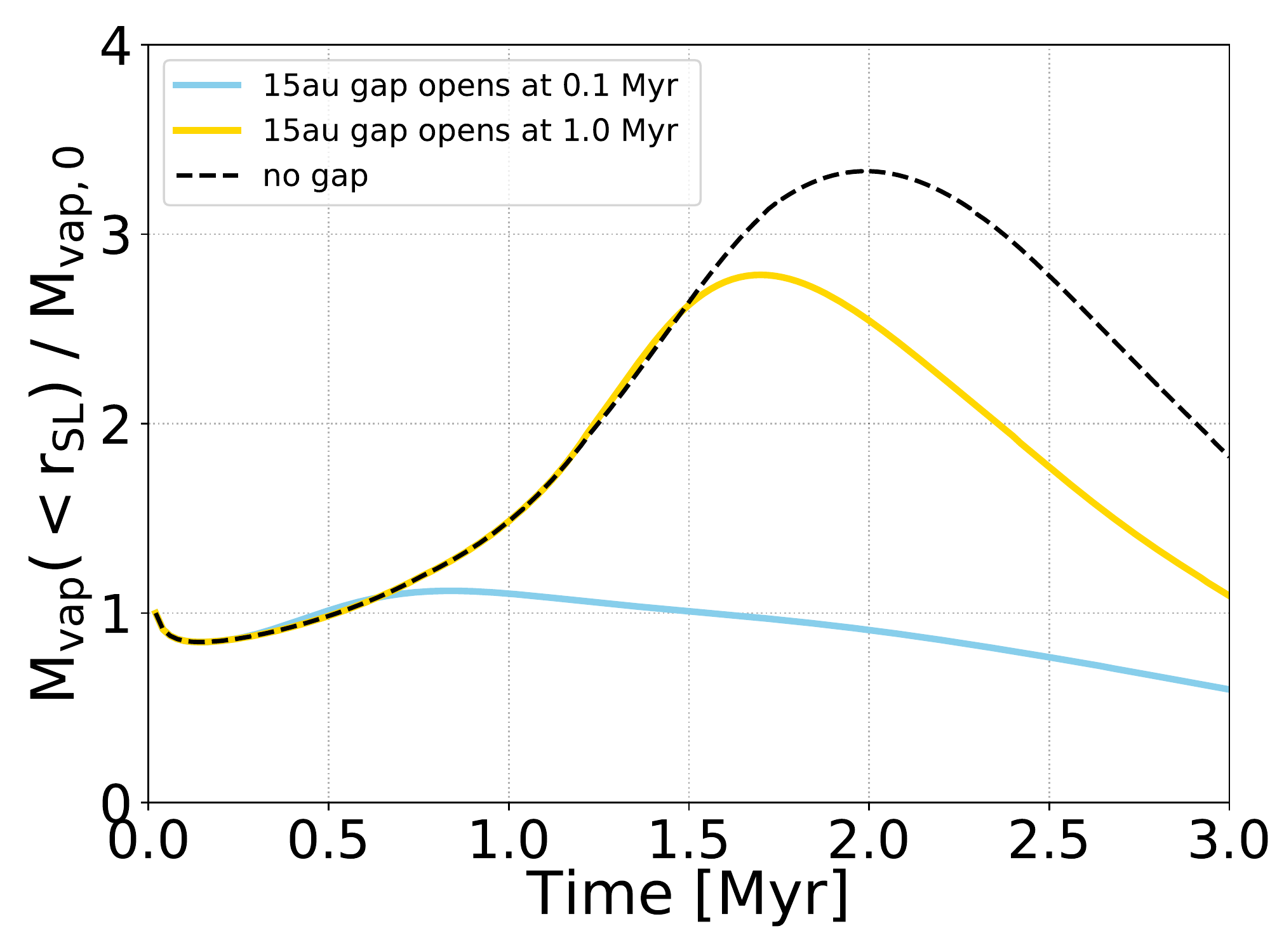}
	\caption{Time evolution of vapor mass (normalized to time t=0) in the inner disk for simulations with 1 mm pebbles and a gap at 15 au, with different gap-opening times (0.1 and 1 Myr). Time evolution for the no-gap case is also shown for comparison.} 
	\label{fig:gapopening}
\end{figure}

We performed one simulation with 1 mm pebbles where the gap at 15 au (i.e., most effective gap) opens at 1 Myr instead of 0.1 Myr (as done for all simulations in this work) and show these results in Figure \ref{fig:gapopening}. We find that as predicted, a later gap-opening allows for higher overall integrated pebble mass flux into the inner disk over 3 Myr. The effect of gap-opening time is dependent on the pebble size, and how fast they drift inward in a disk of our chosen size. For 1 mm pebbles that generally drift inward of the snowline during 1-3 Myr, most pebbles are already within 15 au by 1 Myr. Therefore, the profile of normalized vapor mass with time approaches that of the no-gap case (Figure 3). 

\end{document}